\documentclass[preprint,showpacs,preprintnumbers,nofootinbib]{revtex4-1}
\usepackage{graphicx, color} 
\usepackage{amsmath,amssymb}
\usepackage{ulem} %\sout{hogehoge}
\usepackage{epsf}
\usepackage{color}

\usepackage[italicdiff]{physics}
\usepackage{bm}
\usepackage{mathtools}
\usepackage{enumitem}
\usepackage{wrapfig}
\usepackage{footnote}
\usepackage{here, comment}
\usepackage{bbm}%for bold identity symbol
\usepackage{shorttoc}

\usepackage{subfig}
\usepackage{here, comment}

\newcommand{\eq}[2]{\begin{equation} \label{eq:#1} #2 \end{equation}}%eq with label
\newcommand{\eqn}[1]{\begin{equation} #1 \end{equation}}%eq without labels

\newcommand{\al}[1]{\begin{align} #1\end{align}}

\allowdisplaybreaks[1]

\newcommand{\nn}{\nonumber\\}

\newcommand{\be}{\begin{equation}}
\newcommand{\e}{\end{equation}}
\newcommand{\aln}[1]{\begin{align}#1\end{align}}

\begin{document}
	\count\footins = 1000 %whenever footnote is not working (e.g. overlaps with page number)
	\pagestyle{empty}

	\preprint{KEK-TH-2320}
	
	\title{Non-Gaussianity of  Entanglement Entropy \\  and 
	 Correlations of Composite Operators} 
	\author{Satoshi Iso$^{a,b}$} 
	\email{satoshi.iso@kek.jp}
	\author{Takato Mori$^{a,b}$}
	\email{moritaka@post.kek.jp}
	\author{Katsuta Sakai$^a$}
	\email{sakaika@post.kek.jp}
	
	\affiliation{
		$^a$ KEK Theory Center, High Energy Accelerator Research Organization (KEK), Oho 1-1, Tsukuba, Ibaraki 305-0801, Japan.\\
		$^b$ The Graduate University for Advanced Studies (SOKENDAI), Oho 1-1, Tsukuba, Ibaraki 305-0801, Japan.
	}
	
	\begin{abstract} 
This is an extended version of the previous paper \cite{Iso:2021vrk} to study
entanglement entropy (EE) of a half space in interacting field theories.
In the previous paper,  we have proposed a novel method to calculate EE based on 
the notion of $\mathbb{Z}_M$ gauge theory on Feynman diagrams, and shown that
EE consists of two particular contributions, one from a {\it renormalized two-point correlation function} 
in the two-particle irreducible (2PI) formalism  
and another from {\it  interaction vertices}. 
In this paper, we further investigate them in more general field theories 
and show that the non-Gaussian  contributions from vertices
can be interpreted as renormalized correlation functions of composite operators. 
% Thus EE is being encoded in low-energy correlation functions 
% of various composite operators in a consistent manner with renormalization.  

	\end{abstract}
	%\pacs{??? }
	\maketitle
	\pagestyle{plain}
	\tableofcontents
	%%%%%%%%%%%%%%%%%%
%%%%%%%%%%%%%%%%%%%%%%%%%%%%% Introduction %%%%%%%%%%%%%%%%%%%%%%%%%%%%%%%%%%
\section{Introduction}
\label{s:intro}
%%%%%%%%%%%%%%%%%%%%%%%%%%%%%%%%%%%%%%%%%%%%%%%%%%%%%%%%%%%%%%%%%%%%%%%%%%

	Over the past few decades, entanglement entropy (EE) has attracted much interest in various fields. 
	Originally it is proposed as a nice measure for bipartite entanglement within a pure state and has been widely discussed in the context of quantum information.
However, its usefulness is not limited to quantum information and its usage in physics is very rich and nontrivial. 
In condensed matter physics and lattice quantum field theories, EE is a candidate of order parameters to describe quantum phase transition and topological orders for its nonlocality \cite{PhysRevA.66.032110,Osterloh_2002,PhysRevLett.90.227902,Jin_2004,Holzhey:1994we,Calabrese_2004,PhysRevLett.96.110404,PhysRevLett.96.110405,Ibieta-Jimenez:2019wwo}. In the context of quantum gravity, the Ryu-Takayanagi formula \cite{Ryu:2006bv,Ryu:2006ef,Hubeny:2007xt}, one particular realization of the holography, relates EE of a boundary theory to the extremal surface in the bulk. Furthermore, the information paradox of black holes has been studied in terms of the fine-grained entropy or EE \cite{Solodukhin:2011gn,Almheiri:2020cfm,Almheiri:2019hni,Penington:2019kki}. As EE captures quantum correlation, it is also used to investigate  decoherence in cosmology \cite{Nambu:2008my}.

	Particularly, EE is a useful quantitative measure for correlations between two spatially separated regions in a ground state.
	Despite its significance, practical computations of EE in field theories are not an easy task. 
	If we can exactly diagonalize the reduced density matrix, the EE can be obtained in a straightforward manner, but
	such a diagonalization is not generically possible in  quantum field theories (QFTs). 
Thus, EE has been  intensively discussed in two particular classes of theories,  conformal field theories (CFTs) and Gaussian ({i.e. free}) theories.

For the former class of theories, we can make the most of the conformal symmetry and 
many quantities of interest can be determined by their responses to the conformal transformations \cite{Calabrese_2004,Ruggiero:2018hyl,Hung:2014npa,Casini:2010kt}. In QFTs which can be treated
as perturbations from CFTs, these tools are still available \cite{Rosenhaus:2014zza,Rosenhaus:2014woa,Rosenhaus:2014ula}. 
Alternatively, EE of CFTs with gravity duals is  calculated using holography,
leading to its
geometrical interpretation \cite{Ryu:2006bv,Ryu:2006ef}. (see \cite{Nishioka_2009,RevModPhys.90.035007} for a review).

On the other hand, for the latter class of theories, 
the density matrix of the vacuum is Gaussian and in principle, we can perform an explicit calculation \cite{PhysRevA.70.052329,Katsinis:2017qzh,Bianchi:2019pvv,Lewkowycz:2012qr}. 
There are several ways to deal with it, as summarized for example in \cite{Casini_2009,RevModPhys.90.035007}. 
In the Gaussian case, we can also evaluate EE for a curved boundary or in a curved space by the heat kernel method~\cite{Solodukhin:2011gn,Lewkowycz:2012qr,Hertzberg:2010uv}. 
Fermionic extensions are also studied \cite{Herzog:2013py}.

In comparison to the above two particular classes of theories, 
we have little understanding of EE for general interacting QFTs, 
apart from exactly solvable cases \cite{Donnelly:2019zde}. 
In some supersymmetric theories, the localization method enables an exact calculation of the free energy and EE, 
\cite{Nishioka_2009,Jafferis:2011zi,Pufu:2016zxm,Nishioka:2013haa,RevModPhys.90.035007}. 
EE in interacting theories also 
discussed in perturbative \cite{Hertzberg:2012mn,Chen:2020ild}, nonperturbative \cite{PhysRevB.80.115122,Akers:2015bgh,Cotler:2015zda,Fernandez-Melgarejo:2020utg,Fernandez-Melgarejo:2021ymz,Bhattacharyya:2017pqq} 
or lattice \cite{Wang_2014,Buividovich:2008kq,Buividovich:2008gq,Itou:2015cyu,Rabenstein:2018bri} approaches. 
Nonperturbative studies have taken advantage of the large-$N$ analysis and the renormalization group (RG) flow in the $O(N)$ vector model \cite{PhysRevB.80.115122,Whitsitt:2016irx,Akers:2015bgh,Hampapura:2018uho} or variational trial wave functions \cite{Cotler:2015zda,Fernandez-Melgarejo:2020utg,Fernandez-Melgarejo:2021ymz} or instanton formalism~\cite{Bhattacharyya:2017pqq}. 
These works have partly grasped the behavior of EE relevant to renormalization and beyond free theories. 
Despite these studies, there are many  issues yet to be understood.

Interactions generally bring about two important consequences in QFTs:
renormalization of IR quantities and non-Gaussianity of the vacuum wave function.
Therefore, it is important in the studies of EE to disentangle these two essentially different effects
of interactions. Renormalization is, of course, related to the UV divergences of field theories, but
note that UV divergences of EE are already present in free theories.
It is simply because there are infinitely many microscopic degrees of freedom that contribute
to EE. This type of UV divergences 
should be regularized by suitably renormalizing parameters in the background gravity 
\cite{Cooperman:2013iqr,Barrella:2013wja,Taylor:2016aoi,Taylor:2020uwf,Liu:2012eea,Liu:2013una}. 
In addition,  further renormalizations  are necessary for interacting theories. 
In our previous paper \cite{Iso:2021vrk} we showed 
that the Gaussian part of EE is given by a renormalized two-point correlation function in the
two-particle irreducible (2PI) formalism \cite{PhysRevD.10.2428,Berges:2004yj} 
and thus UV divergences specific to interactions are appropriately taken into account.
But it is not the end of the story of EE in interacting theories. 
Besides renormalization of IR quantities, interactions induce  non-Gaussianity of the vacuum wave function
and we wonder how much 
EE is affected by such non-Gaussianity. 
Non-Gaussiainity of the vacuum can be expressed in terms of the connected multi-point correlation functions
(due to the Wick theorem), and thus the non-Gaussian contributions to EE must be 
written in terms of  higher point Green functions. 
 In the analysis of the  $\phi^4$ scalar theory in  \cite{Iso:2021vrk}, 
by extracting the  Gaussian contributions to EE in 2PI formalism, we have 
uncovered purely non-Gaussian contributions from four-point vertex functions.

In this paper, we continue the investigations of EE proposed in  the previous paper  \cite{Iso:2021vrk}
and study further issues of renormalization and non-Gaussianity in a general field-theoretical approach.
Especially we will show that the purely non-Gaussian contributions associated with the four-point vertex
functions can be interpreted as contributions from 
renormalized two-point correlation functions of {\it composite operators.} 
The result indicates that all the non-Gaussian contributions to 
 EE can be interpreted as a sum of contributions from 
renormalized two-point correlation functions of various  composite operators. 
We also show that the analysis is not restricted to scalar field theories but applicable 
to  general QFTs with nonzero spins.

The paper is organized as follows. 
In Section \ref{s:orbifold}, we give a brief review of the orbifold method. 
The calculation of EE in a free field theory is demonstrated and a generalization to higher spin fields is commented. 
 In Section \ref{s:gaugetheoryF}, we lift the method to interacting cases and provide 
 a general  methodology, $\mathbb{Z}_M$ gauge theory on Feynman diagrams, to compute the free energy and EE. 
 We then prove the area law of EE for interacting cases in the framework. %There
Section \ref{s:prop} is devoted to the analysis of contributions to EE from propagators
with more detailed explanations and discussions on the calculations in~\cite{Iso:2021vrk}. 
%It is a more detailed and rigid discussion for our previous work~\cite{Iso:2021vrk}. 
We introduce the two-particle irreducible (2PI) formalism and express the Gaussian part of EE in terms of the renormalized propagator. 
Next, we move to the investigation of the non-Gaussian contributions in Section \ref{s:vert}. 
We see that interaction vertices have contributions as well as 
the Gaussian contributions. 
We show that these non-Gaussian contributions can be understood  as renormalized
two-point correlations in terms of  the corresponding composite operators. 
We also discuss an extension  to a general theory with spins in Section \ref{s:spin}. 
Finally, we make a summary and discussion towards future studies in Section \ref{s:discussion}.

%%%%%%%%%%%%%%%%%%%%%%%%%%%%%%%%%%%%%%%
\section{Replica trick and orbifold method}
\label{s:orbifold}
\begin{figure}[t]
	\centering
	\includegraphics[width=10cm,clip]{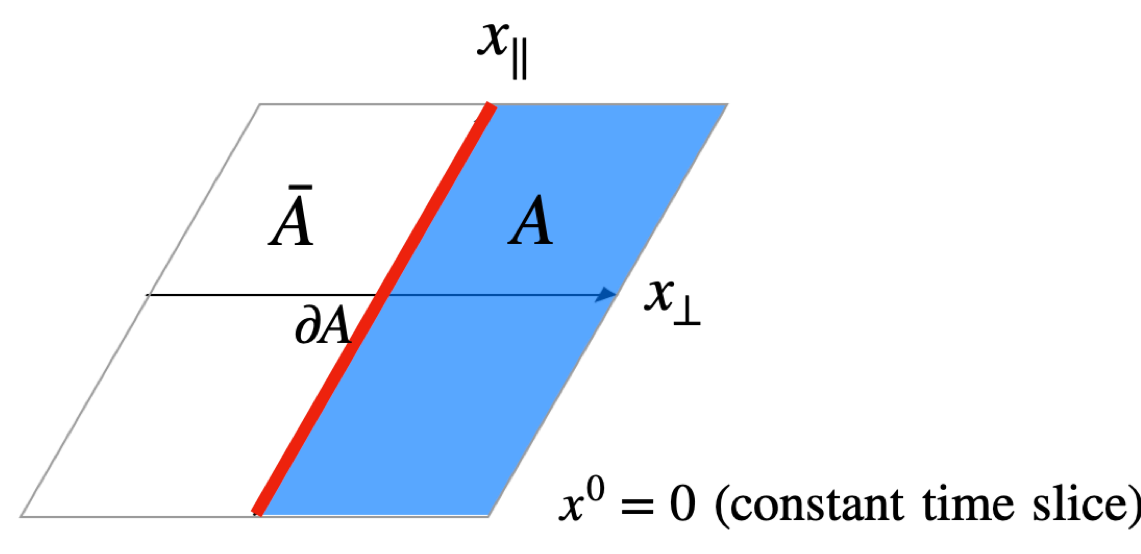}
	\caption{Our choice of the subregion $A$ and its complement $\bar{A}$. It is a half space given by $A=\{x^0=0,x_\perp\ge 0, \forall x_\parallel\}$. The boundary of the subregion is parametrized as $\partial A=\{x^0=0,x_\perp=0,\forall x_\parallel\}$.}
	\label{fig:subregion}
\end{figure}

%%%%%%%%%%%%%%%%%%%%%%%%%%%%%%%%%%%%%%%
First, we review the replica trick and the orbifold method to calculate EE. % \cite{Nishioka_2007,He:2014gva}. 
Consider a Hilbert space which consists of two subspaces 
corresponding to the physical subsystems of interest $A$ and $\bar{A}$: 
$\mathcal{H}_{\mathrm{tot}}=\mathcal{H}_A\otimes\mathcal{H}_{\bar{A}}$. 
The EE of $A$ is defined as 
\[ S_{A}=-\Tr_{A}\rho_{A}\log\rho_{A},\]
where $\rho_A=\Tr_{\bar{A}}\rho_\mathrm{tot}$ is a reduced density matrix of 
the total one, $\rho_\mathrm{tot}$. 
One of the standard methods to calculate EE in QFTs 
is known as the replica method \cite{Holzhey:1994we,Calabrese_2004}:
\aln{
	S_A%=-\Tr_A \left(\rho_A\log \rho_A \right)
	:=\lim_{n\rightarrow 1}\left[\frac{1}{1-n}\log \Tr\rho_A^n\right]
	= -\lim_{n\rightarrow 1} \pdv{n} \left[\Tr_A\rho_A^n\right].
	\label{eq:replica}
}
Note that for EE to be uniquely determined by the replica trick, we assume the analytical continuation of $n\in\mathbb{Z}_{>1}$ to $\mathbb{R}$. This formula holds for general $A$ and $\rho$ as long as such an analytical continuation exists. 
In this paper, we take $A$ as a half space on a time slice in a $(d+1)$-dimensional  spacetime,
$A=\{x^0=0,x_\perp\ge 0, \forall x_\parallel\}$, where $x_0$ is a Lorentzian temporal coordinate while 
$x_\perp$  is a one-dimensional normal direction and $x_\parallel$ are the rest $(d-1)$-dimensional
parallel directions %directions 
to $\partial A$ %respectively 
(Fig.\ref{fig:subregion}).
%%%%%%%%%%%%%%%%%%%%%%
%%%%%%%%%%%%%%%%%%%%%%

%%%%%%%%%%%%%%%%%%%%%%
\begin{figure}[t]
	\begin{tabular}{c}%prevent line break
		\hspace*{-0.05\linewidth}
		\begin{minipage}{0.5\hsize}%align figs horizontally
			\centering
			\includegraphics[width=\linewidth]{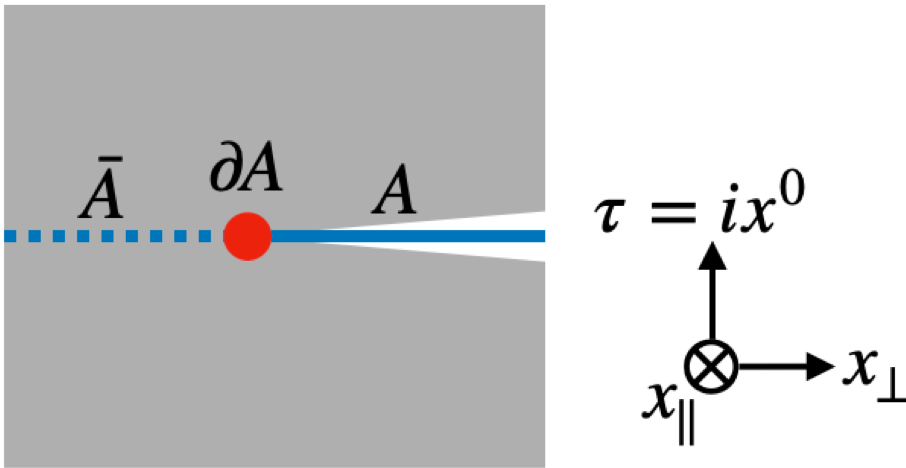}
		\end{minipage}
		\hspace*{0.05\linewidth}
		\begin{minipage}{0.3\hsize}
			\centering
			\includegraphics[width=\linewidth]{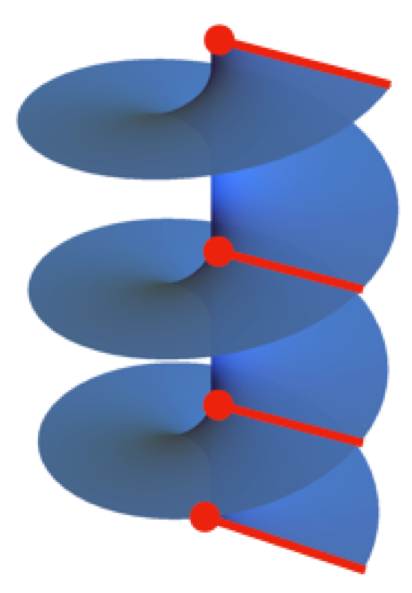}
		\end{minipage}
	\end{tabular}
	\caption{
		The Euclidean path integral representation of our reduced density matrix $\tilde{\rho}_A$ (left) and its $n$-fold cover $\tilde{\rho}_A^n$ ($n=3$) (right).
		}
	\label{fig:euclid-pathint}
\end{figure}
%%%%%%%%%%%%%%%%%%%%%%

Let us define an unnormalized density matrix $\tilde{\rho}_\mathrm{tot}$ such that
$\rho_\mathrm{tot}:=\tilde{\rho}_\mathrm{tot}/Z_1$, 
%\[\rho_\mathrm{tot}=\frac{\tilde{\rho}_\mathrm{tot}}{Z_1},\]
where $Z_1$ is a partition function of the total system on $\mathbb{R}^2\times\mathbb{R}^{d-1}$, 
where $\mathbb{R}^2$ is spanned by two normal coordinates, the Euclidean time $\tau=ix^0$ and $x_\perp$, with respect to $\partial A$ and the rest $\mathbb{R}^{d-1}$ is spanned by $x_\parallel$. Then, with an unnormalized reduced density matrix defined by $\tilde{\rho}_A := \Tr_{A} \tilde{\rho}_\mathrm{tot}$, $\Tr \tilde{\rho}_A^n$ is regarded as a partition function of the theory on $\Sigma_n\times\mathbb{R}^{d-1}$, where 
$\Sigma_n$ is an $n$-folded cover of a two-dimensional plane spanned by  %the Euclidean time 
$\tau$ and $x_\perp$ %, and thus, 
or equivalently a two-dimensional cone with a deficit angle $2\pi(1-n)$.
The Euclidean path integral representation of the reduced density matrix and the replicated one is shown in Fig.\ref{fig:euclid-pathint}. The EE can be rewritten as 
\begin{equation}
	S_A=  \left.\pdv{F_n}{n} \right|_{n\to1}-F_1
	\label{e:EE_n}
\end{equation}
 in terms of the free energy  on $\Sigma_n \times \mathbb{R}^{d-1}$
\[
F_n:=-\log \Tr_A\tilde{\rho}_A^{\,n}.
\] 
%Here $\tilde{\rho}_A$ is not necessarily normalized 
%and $\rho_A= \tilde{\rho}_A/Z_n$.

\vspace{5mm}
We can further proceed to reduce the calculation employing the orbifold method~\cite{Nishioka_2007,He:2014gva}. In this method, we analytically continue $n$ to $1/M$ with an integer $M$ to obtain the theory on the orbifold $\mathbb{R}^2/\mathbb{Z}_M\times \mathbb{R}^{d-1}$ instead of a cone. Eq.(\ref{e:EE_n}) is then rewritten in terms of the free energy $F^{(M)}=F_{1/n}$ as
\aln{
	S_A=-\frac{\partial \left(M F^{(M)}\right) }{\partial M}\bigg|_{M\to 1},
	\label{eq:EE_M}
}
provided that $M\in\mathbb{Z}_{>1}$ can be analytically continued to 1.
A state on the orbifold can be obtained by acting the $\mathbb{Z}_M$ projection operator \cite{Gersdorff_2008}, 
\[
\hat{P}=\sum_{m=0}^{M-1}\frac{ \hat{g}^{\, m} }{M},
\]
on a state in an ordinary  flat plane, 
where $\hat{g}$ is a $2\pi/M$ rotation operator {around the origin},
\aln{
\hat{g}\left|x,\bar{x},x_\parallel\right>=\left|e^{2\pi i/M}x,e^{-2\pi i/M}\bar{x},x_\parallel\right>.
} 
In the following discussion, we will call this $\mathbb{Z}_M$ rotation 
 $\hat{g}^m$ as an  $m$-\textit{twist}, where $ m \in \mathbb{Z}\mod M$. 
%Here 
$\bm{x}=(x,\bar{x})$ are complex coordinates for the perpendicular directions, $x=x_\perp+i\tau,\, \bar{x}=x_\perp-i\tau$. 
%Hereafter they are denoted by $\bm{x}$. 
%In order to perform the summation over $m$ from $1$ to $M-1$, %in a way such that $M$ is analytically continued to $\mathbb{R}$, 
%it is essential to turn it into a complex integral, picking up the residues $\{e^{2\pi i m/M}\}_{m=1, \cdots, M-1}$ \cite{doi:10.1063/1.531345}. By this treatment, the analytic continuation is unique due to the Carlson's theorem \cite{Casini_2009,Witten:2018xfj}.

\medskip
By using the orbifold method, EE can be easily calculated for free  theories \cite{Nishioka_2007}. 
In the case of a real scalar field theory, the free energy takes the following form,
\aln{
F_\text{free}^{(M)}&=\frac{1}{2}\mathrm{Tr}\,\mathrm{log}[\hat{P}(-\Box+m^2_{0})]\\
&=\frac{1}{2}\int\frac{d^2\bm{k}}{(2\pi)^2}\frac{d^{d-1}k_\parallel}{(2\pi)^{d-1}}\mathrm{log}(k^2+m_{0}^2)\left<\bm{k},k_\parallel\right|\hat{P}\left|\bm{k},k_\parallel\right>. 
\label{eq:free-energy-free-scalar}
}
The trace of $\hat{P}$ is computed as follows.
The diagonal matrix element of $\hat{g}^m$ is given by
\aln{
		\ev{\hat{g}^m}{\bm{k},k_\parallel}&=\bra{{k}_\parallel}\ket{{k}_\parallel}\ev{\hat{g}^m}{k,\bar{k}} %\nonumber\\
	%	&=V_{d-1}\bra{\omega^{-m}k,\omega^m \bar{k}}\ket{ k,\bar{k}}
	%	\nonumber\\
         =(2\pi)^2V_{d-1}\delta(\omega^mk-k)\delta(\omega^{-m}\bar{k}-\bar{k}) ,
   %      \nonumber
     }
where $V_{d-1}=(2\pi)^{d-1} \delta^{d-1}(k_\parallel-k_\parallel)$ is a transverse $(d-1)$-dimensional volume 
 and $\omega=e^{2\pi i/M}$.
For $m\neq 0$, this becomes
 \aln{
		\ev{\hat{g}^m}{\bm{k},k_\parallel}=(2\pi)^2 V_{d-1} \delta^{2}(\bm{k}) \frac{1}{\omega^m-1}\frac{1}{\omega^{-m}-1}.
		\label{eq:twist-k}}
For $m\neq0$, it is proportional to $V_{d-1}$, the area of the boundary $\partial A$. On the other hand, for
$m=0$, it is proportional to 
$V_{d-1}\times\delta^2(\bm{0})\propto V_{d+1}$, the volume of the 
whole region of the path integral. % $A$
From this, we  see that  twisting a propagator with $m \neq 0$ 
constrains the normal components of the momentum zero, ${\bm k}=0$.
In Sec.\ref{sec:twist}, we will have a slightly different interpretation of the twisted propagator. 

\begin{figure}
	\centering
	\includegraphics[width=7cm,clip]{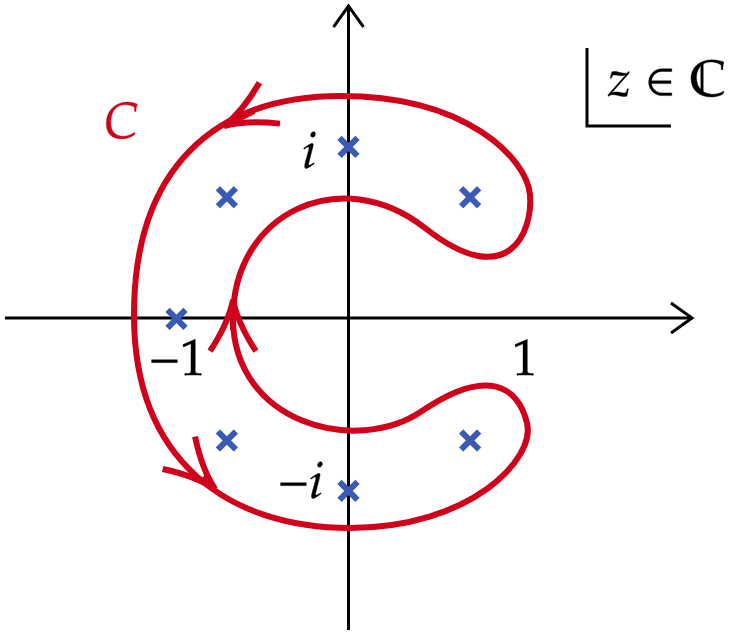}
	\caption{The contour $C$ (red curve) and simple poles of $p(z)$ (blue cross marks) in Eq.\eqref{eq:complex-int} ($M=8$, as an example).}
	\label{fig:contour}
\end{figure}

The summation over $m$ from $1$ to $M-1$ can be performed as follows.
Given a holomorphic function $f(z)$, its summation is given by
\eq{complex-int}{\sum_{m=1}^{M-1}f\left(\omega^{m}\right)=\oint_C \frac{\dd{z}}{2\pi i} p(z) f(z),}
where
\eqn{p(z)\equiv \frac{Mz^{M-1}}{z^M -1}-\frac{1}{z-1}}
has simple poles at $\omega^m =e^{2\pi m  i/M} $ with $m=1, \cdots, M-1$. The integration contour $C$ is chosen to surround these poles \cite{doi:10.1063/1.531345} (Fig.\ref{fig:contour}).
Then the summation of Eq.\eqref{eq:twist-k} is calculated as
\al{
	\sum_{m=1}^{M-1} \frac{1}{\omega^m-1}\frac{1}{\omega^{-m}-1}
	 %&    =\oint_C \frac{\dd{z}}{2\pi i} p(z) \frac{1}{z-1} \frac{1}{z^{-1}-1}\nonumber\\
	=\oint_{C_1}\frac{\dd{z}}{2\pi i} \frac{z p(z)}{(z-1)^2} 
%	&=\left.\pdv{z}\left(z p(z)\right)\right\vert_{z=1}\nonumber\\
	=\frac{M^2-1}{12},
	\label{eq:twist-sum}}
where $C_1$ is a counter-clockwise circle around $z=1$. 
It is written as a compact form,
\aln{
\sum_{m=1}^{M-1}\frac{1}{\sin^2\left(\frac{\pi m}{M}\right)}=\frac{M^2-1}{3}.
\label{e:summ}
}

\vspace{5mm}
Plugging Eq.\eqref{eq:twist-sum} into Eq.\eqref{eq:free-energy-free-scalar}, we obtain
%Thus, we have
\aln{
F_{\text{free}}^{(M)} 
&= \frac{1 }{2 M} \int \frac{d^2 \bm{k} \ d^{d-1}k_{\parallel}} {(2 \pi)^{d-1}}   \log (k ^2 + m^2_{0}) \left( \frac{V_{d+1}}{(2\pi)^2}+  V_{d-1}\frac{M^2-1}{12}  \delta^2(\bm{k}) \right) .
\label{eq:1-loop-F}
}
The first term  proportional to $V_{d+1}$ vanishes in the calculation of EE in Eq.(\ref{eq:EE_M}). 
We will see later that such property generally holds even in presence of interactions. 
On the other hand, %we have 
the second term is proportional to the area $V_{d-1}$ survives in Eq.(\ref{eq:EE_M})
due to an additional $M$-dependence of  $M^2 -1$. 
Consequently, 
EE for a free scalar theory is given by
\aln{
S_{\text{1-loop}}= - \frac{V_{d-1} }{12}  \int^{1/\epsilon} \frac{d^{d-1}k_{\parallel}} {(2 \pi)^{d-1}}   
\log \left[ (k_{\parallel} ^2 + m^2_{0}) \epsilon^2 \right].
\label{EE-1loop}
}
%which obeys the area law of EE. %Here we have introduced a UV cutoff scale $\epsilon$ is introduced. 
Here we have introduced a UV cutoff $\epsilon$, which is naturally identified as a lattice spacing for a lattice system.
Note that the EE decreases as the mass increases.

The calculation for the scalar theory can be easily generalized to bosonic higher spin theories \cite{He:2014gva}. 
In this case, a state is parametrized by $|{\bm x}, x_\parallel; s \rangle$, where $s$ is a spin of $SO(2)$ rotation. 
Then the action of the two-dimensional rotation $\hat{g}$ is given by
\aln{
\hat{g}\left|x,\bar{x},x_\parallel;s \right>= e^{2\pi s i/M }
\left|e^{2\pi i/M}x,e^{-2\pi i/M}\bar{x},x_\parallel; s \right>
} 
and the sum over $m \neq 0$ in Eq.(\ref{e:summ})
 is replaced by 
\aln{
\sum_{m=1}^{M-1}\frac{ \cos (\frac{2\pi ms}{M} ) }{\sin^2\left(\frac{\pi m}{M}\right)}=
\frac{1}{3} \left[M^2-1+6M^2\left( \left\{ s/M\right\}^2 -  \left\{ s/M\right\}\right) \right],
\label{e:summ-highers}
} 
where $\{ x \}$ is a fractional part of $x$. 
For fermionic generalizations, we need special care since $2\pi$ rotation
gives an extra minus sign, $\hat{g}^M=-1$, and it cannot be regarded as $\mathbb{Z}_M$ orbifold. 
To overcome this difficulty, the authors in  \cite{He:2014gva} take  an odd $M$ and consider 
$\hat{g}^2$ as the generator of $\mathbb{Z}_M$ orbifold on a double cover of the Riemann surface. 
Another subtlety  in higher spin generalizations
in  analytical continuation of $M$ since Eq.(\ref{e:summ-highers}) contains a
non-analytic function, $\left\{ s/M\right\}$, and we need to constrain the value of $s$ within $[-M, M]$ for fermions
or $[0,2M]$ for bosons. Thus the calculation of EE for higher spins than 3/2 may have subtlety in the orbifold method. 
For more details, see  \cite{He:2014gva}. 

%%%%%%%%%%%%%%%%%%%%%%%%%%%%%%%%%%%%%%%%%%%%%%%%
\section{Orbifold method in an interacting field theory}
\label{s:gaugetheoryF}
%%%%%%%%%%%%%%%%%%%%%%%%%%%%%%%%%%%%%%%%%%%%%%%%
In this section, we apply the orbifold method to interacting field theories and calculate the free energy
of the $\mathbb{Z}_M$ orbifold.  Each propagator in a Feynman diagram is projected by the projection
operator $\hat{P}$ and thus we need to sum all the twists, $m \in {\bf Z}_M$ in each propagator. 
But  there are redundancies in the summation, 
associated with  $\mathbb{Z}_M$ rotations at each vertex in Feynman diagrams, 
 and it is  not so trivial to extract relevant terms
that  contribute to EE in  Eq.(\ref{eq:EE_M}). 
We first show that such redundancies can be 
systematically treated by  performing the summation in the framework of  the
 $\mathbb{Z}_M$ gauge theory on Feynman diagrams.
 Namely,  assign  $\mathbb{Z}_M$ twists on  each link (i.e., on a propagator) and  
 define  $\mathbb{Z}_M$ gauge transformations  on each vertex, and take a summation
 over all the twists modulo $\mathbb{Z}_M$ gauge transformations. 
 Then, a gauge-invariant  configuration of twists  is characterized 
by a set of fluxes of twists on each plaquette of each Feynman diagram. 
Within this framework, we can easily prove  the area law of  EE.

\subsection{Setup}
Consider, for simplicity, a $\phi^4$ scalar field theory on  a $\mathbb{Z}_M$ orbifold without a non-minimal coupling
to the curvature.   
The action is given by 
\aln{
I=	\int_{\mathbb{R}^2} \frac{d^2x}{M} \int_{\mathbb{R}^{d-1}} d^{d-1}x_\parallel \left[\frac{1}{2}
%\phi \hat{P} 
(\hat{P} \phi) \left(-\Box+m^2_{0} \right)
(\hat{P}\phi )
+\frac{\lambda}{4}(\hat{P}(\phi))^4\right]
	\label{eq:action} 
}
in terms of a field $\phi(x)$ in flat space $\mathbb{R}^2\times\mathbb{R}^{d-1}$ %and 
but with the projection operator,
where
\aln{
\hat{P}\phi(\bm{x},x_\parallel):=\frac{1}{M}\sum_{m=0}^{M-1}\phi(\hat{g}^m\bm{x},x_\parallel).
}%--------------------------
From the action Eq.(\ref{eq:action}), 
the inverse propagator of the orbifold theory in flat space is given by
\aln{
\hat{G}_0^{-1\, (M)}= \frac{\hat{P}\hat{G}_0^{-1}\hat{P}}{M} =  \frac{\hat{P} \left(-\Box+m^2_{0} \right)   \hat{P}}{M} ,
\label{propagator-orbifold}
}
%Using the relation
%\[
%\hat{G}_0^{-1\, (M)} \frac{1}{M} \hat{G}_0^{(M)}=\hat{G}_0^{(M)} \frac{1}{M} \hat{G}_0^{-1\, (M)}=\hat{P},
%\]
and  the propagator, which satisfies $\hat{G}_0^{-1\, (M)} \hat{G}_{0}^{(M)}=\hat{P} $, is then written as
\aln{
G_{0}^{(M)} (x,y) &= M\bra{x}(\hat{P} \hat{G}_0 \hat{P})\ket{y} 
=\sum_{m=0}^{M-1} G_0(\hat{g}^m x,y) ,
% \nonumber \\
%&=  \sum_{n=0}^{M-1} 
%\nonumber
\label{eq:green}
}
where 
\aln{ 
G_0(\hat{g}^m x, y) %= \bra{\hat{g}^n x} G_0 \ket{y} 
&=  \int\frac{d^{d+1} p}{(2\pi)^{d+1}}\frac{e^{i p \cdot(\hat{g}^m x-y)}}{p^2+m^2_{0}} %\nonumber\\&
= \int\frac{d^2\bm{p}}{(2\pi)^2}\frac{d^{d+1} p_\parallel}{(2\pi)^{d+1}}\frac{e^{i \bm{p} \cdot(\hat{g}^m \bm{x}-\bm{y})+ip_\parallel\cdot(x_\parallel-y_\parallel)}}{p^2+m^2_{0}}.
}
The $\mathbb{Z}_M$ rotation on $y$ 
has been eliminated since a projection operator $\hat{P}$ commutes with $\hat{G}_0$.
From the identity ${\bm p}\cdot \hat{g}^m {\bm x}=\hat{g}^{-m} {\bm p} \cdot {\bm x}$, we see that
 the flow-in momentum from the propagator at a vertex $x$ is given by the twisted momentum $(\hat{g}^{-m} {\bm p}, \ p_\parallel)$. 
 In the momentum space representation, the propagator is written as 
\aln{
\left<\bm{p},p_\parallel\right|G^{(M)}\left|\bm{q},q_\parallel\right>=\sum_{m=0}^{M-1}\frac{1}{p^2+m^2_{0}}(2\pi)^{d+1}\delta^2(\hat{g}^m\bm{p}-\bm{q})\delta^{d-1}(p_\parallel-q_\parallel)
}
with $-m$ redefined as $m$.
%On the other hand, 
The interaction vertex in the Euclidean $\phi^4$-theory 
read off from the action Eq.\eqref{eq:action} is
\aln{
-3!\frac{\lambda}{M}.
}
%This is 
The $x$ integration gives 
the ordinary momentum conserving delta functions
 \aln{ \delta^2 \left(\sum_i \hat{g}^{m_i} \bm{p}_i \right)\ \delta^{d-1} \left(\sum_i p_{i\parallel} \right)
 \label{momentum-conserving-deltafuntion}
 }
  with a twisted flow-in momentum.
%Note that the projection operators 
%in the interacting term in the action Eq.(\ref{eq:action} )
%can be absorbed into those in the propagators by using the relation $\hat{P}^2=\hat{P}$.

\medskip

\subsection{Area law of EE in orbifold method} 
\begin{figure}
	\centering
	\subfloat{\includegraphics[width=7cm,clip]{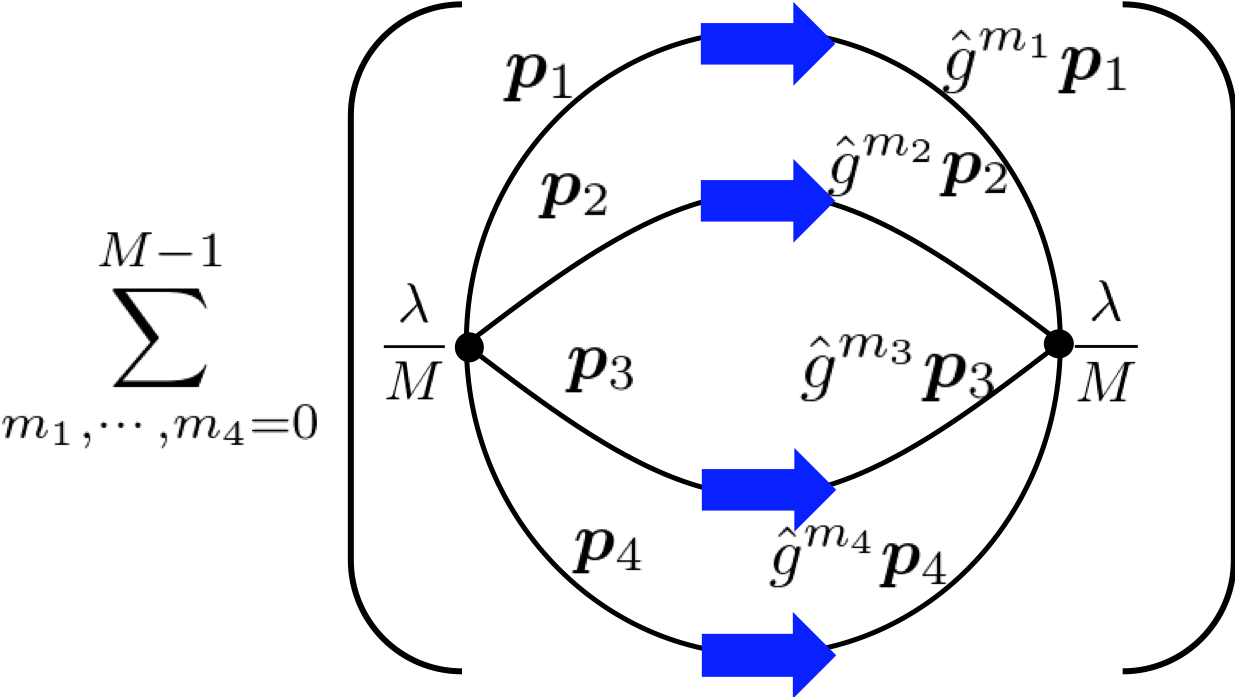}\label{fig:bubble1}}
	\hspace{1cm}
	\subfloat{\includegraphics[width=7cm,clip]{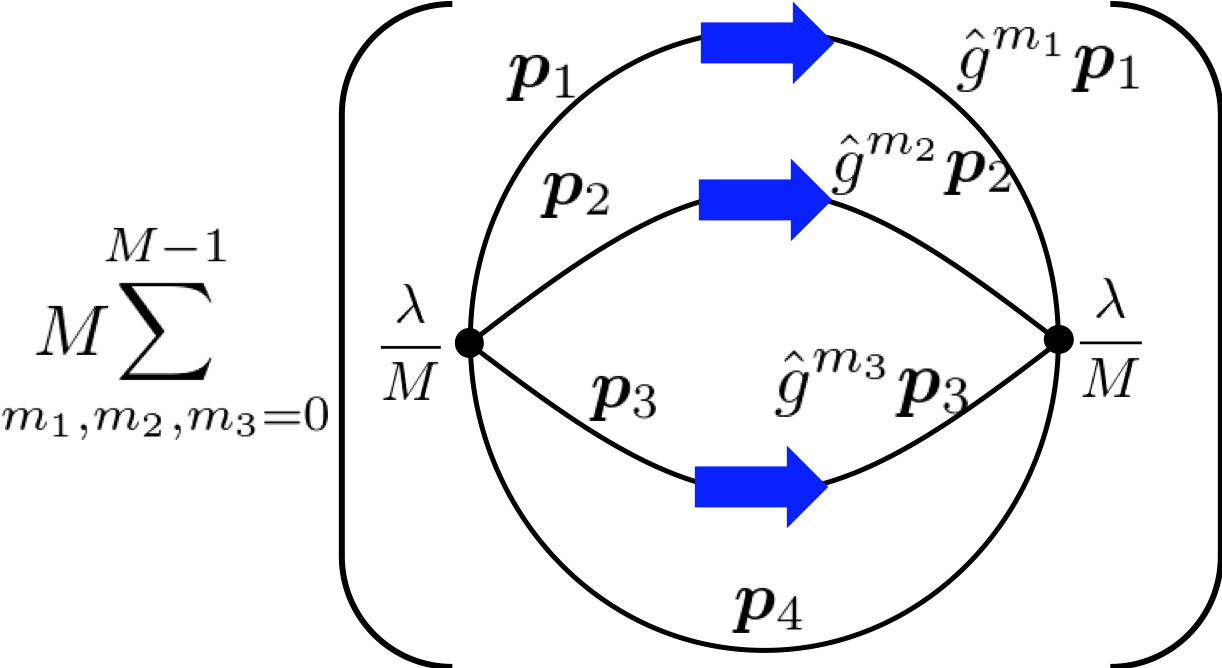}\label{fig:bubble2}}
	\caption{Two equivalent configurations of twists in the three-loop diagram. 
	%Red dashed lines denote the propagators with the momentum twisted. 
	Blue arrows denote twisted momenta with a twist $m_i$. 
The bottom propagator in the right is made untwsited by a $\mathbb{Z}_M$ rotation at a vertex.}
	\label{fig:phi43loops}
\end{figure}

We first show the area law of EE, i.e. $S_A\propto V_{d-1}$ (for a review of this property see \cite{Eisert:2008ur} for example). 
There are two %important points 
factors  responsible for the area law:
%in focusing on a bubble diagram: 
an overall dependence of the free energy on $M$, 
and the nontrivial %volume
argument in the momentum-conserving delta function in Eq.(\ref{momentum-conserving-deltafuntion}). %factor. 
Consider a Feynman diagram with $N_V$ vertices, $N_P$ propagators, and $L$ loops. 
Each vertex has a factor $1/M$, and we may think naively that we have an overall factor of $(1/M)^{N_V}$
in the free energy. 
However, this is not correct for the following reason. While we have $N_P$ summations over the twists, 
some of the $\mathbb{Z}_M$ summations 
are trivial due to the $\mathbb{Z}_M$ invariance of each vertex
\aln{
\delta^2 \left(\sum_i\bm{p}_i \right)=\delta^2 \left(\hat{g}^n\sum_i\bm{p}_i \right)~~~(\hat{g}^n: \text{an arbitrary twist}).%.
}
and give additional overall $M$ factors.
Diagrammatically, we can untwist a part of twisted momenta on the propagators, i.e. eliminate some of the twists 
by using the above invariance of the $\delta$-function.  
Then  the summations over the corresponding twists give  overall factors of $M$.
See Fig.\ref{fig:phi43loops} for an example. 
%%%%%%%%%%%%%%%%%%%%%%%%%%%%%%%%%%
%%%%%%%%%%%%%%%%%%%%%%%%%%%%%%%%%%
However, not all the $\mathbb{Z}_M$ rotations 
at  $N_V$ vertices are independent. 
When we eliminate the twists of momenta by delta functions at vertices, we necessarily encounter the last delta function %who has
with no room for untwisting,
\aln{
\delta^2 \left(\sum_{l=1}^L (1-\hat{g}^{m_l}) \bm{p}_l \right)  \delta^{d-1}(0).
\label{e:residualdelta}
}
Here $p_l$'s are all independent, nothing but the loop momenta. 
The number of the residual twists, which cannot be untwisted anymore, consistently coincides with $L$ because we can untwist %as a result 
$N_V-1$ out of $N_P$ twists.\footnote{Note that $N_V-N_P+L=1$.} %At the same time,
As a result, the trivial sums give an overall factor $M^{N_V-1}$. %Thus,
After all, the overall $M$-dependent factor %$M$-dependence 
of a general bubble diagram is given by
\aln{
\left(\frac{1}{M}\right)^{N_V}\times M^{N_V-1}=\frac{1}{M}.
\label{e:overallM}
}

\vspace{5mm}
As for the second point, we need to look into the argument of the delta functions.
 The $(d-1)$-dimensional delta function in Eq.(\ref{e:residualdelta}) yields the $(d-1)$-dimensional volume: $\delta^{d-1}(0)\propto V_{d-1}$.
Consequently, the diagram itself is formally expressed as
\aln{
\sum_{\{m\}}\frac{1}{M}V_{d-1}\int\prod_{l=1}^L\left[\frac{d^2\bm{p}_l}{(2\pi)^2}\right]I(\{\bm{p}\};\{m\})
\delta^2 \left(\sum_{l=1}^L (1-\hat{g}^{m_l}) \bm{p}_l \right),%.
}
where  
$I(\{\bm{p}\};\{m\})$ is a function of momenta and twists.
We have to sum up diagrams over various configurations of the twists. First, note that the two-dimensional delta function generically has a nontrivial argument. They are proportional to $V_{d-1}$, the area of the boundary. The only exception is the configuration with trivial twists $m_1=\cdots=m_L=0$. It is identical to the corresponding diagram in flat space, where the diagram has a factor 
$\delta^2(\bm{0})$. It is then proportional to $V_{d-1}\times V_2$, the volume of the bulk.

The above statement holds for every bubble diagram. It leads to the following formal expression of the free energy:
\aln{
F^{(M)}=\frac{1}{M}F_\text{flat}+\tilde{F}^{(M)}_\text{twisted},~~~\tilde{F}^{(M)}_\text{twisted}=V_{d-1}\frac{f(M)}{M},
}
where $F_{\text{flat}}$ is the free energy of the $M=1$ field theory and 
$f(M)$ is an intensive quantity with a nontrivial dependence on $M$. 
Now we can readily check the area law from the above expression just in the same manner as in the free theory case. Although the first term is proportional to the volume of the bulk, it does not contribute to EE by the formula Eq.(\ref{eq:EE_M}). In contrast, $\tilde{F}^{(M)}_\text{twisted}$ is proportional to the area of the boundary, and it does contribute to EE due to the $M$-dependence of $f(M)$.
This completes the proof of the area law to all orders. 
The proof applies to any locally interacting theories.\footnote{We have additional phase factors due to spins, but they do not alter the result.}
As a comment, it is also interesting to see a deviation from the area law by applying our formalism to theories exhibiting the volume law (e.g. manifestly nonlocal theories~\cite{Shiba:2013jja} and ones with nonlocal properties in some limit, Lifshitz field theories~\cite{He:2017wla,MohammadiMozaffar:2017nri,Gentle:2017ywk}, for example) or the logarithmic violation to the area law (e.g. (non-)Fermi liquid theories~\cite{Ogawa:2011bz}).

%%%%%%%%%%%%%%%%%%%%%%%%%%%%%%%%%%%%%%%%%%

\vspace{3mm}
\subsection{$\mathbb{Z}_M$ gauge theory on Feynman diagrams}
The above statement is based on the idea that we can eliminate the redundant twists by using the invariance of the vertices under
$\mathbb{Z}_M$ rotations. 
This procedure reminds us of gauge fixing in an ordinary gauge theory. In the following, we will show that this analogy works well in the investigation and that we can extract independent twists in a covariant manner. We call this methodology %the structure the 
$\mathbb{Z}_M$ gauge theory on Feynman diagrams. 
%%%%%%%%%%%%%%%%%%%%%%%%%%%%%%%%%%%%%%%%
\begin{figure}[t]
	\begin{tabular}{c}%prevent line break
		\hspace*{-0.05\linewidth}
		\begin{minipage}{0.35\hsize}%align figs horizontally
			\centering
			\includegraphics[width=\linewidth]{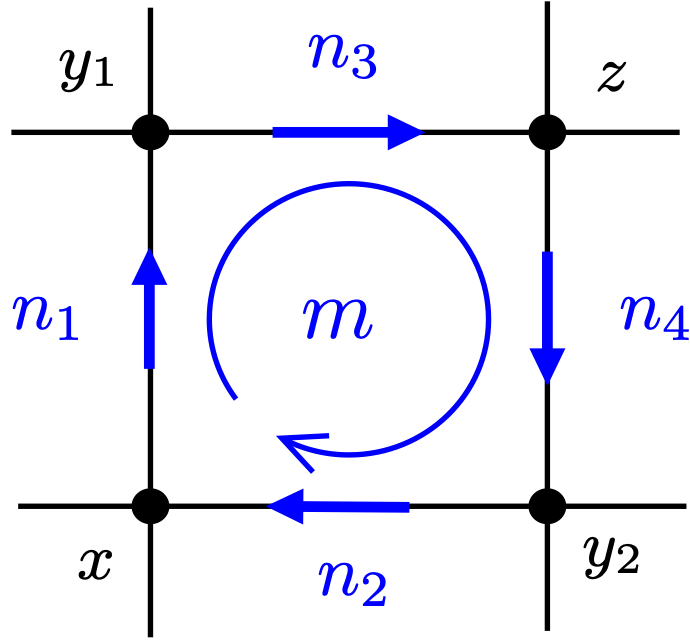}
		\end{minipage}
	\hspace*{0.05\linewidth}
		\begin{minipage}{0.45\hsize}
			\centering
			\includegraphics[width=\linewidth]{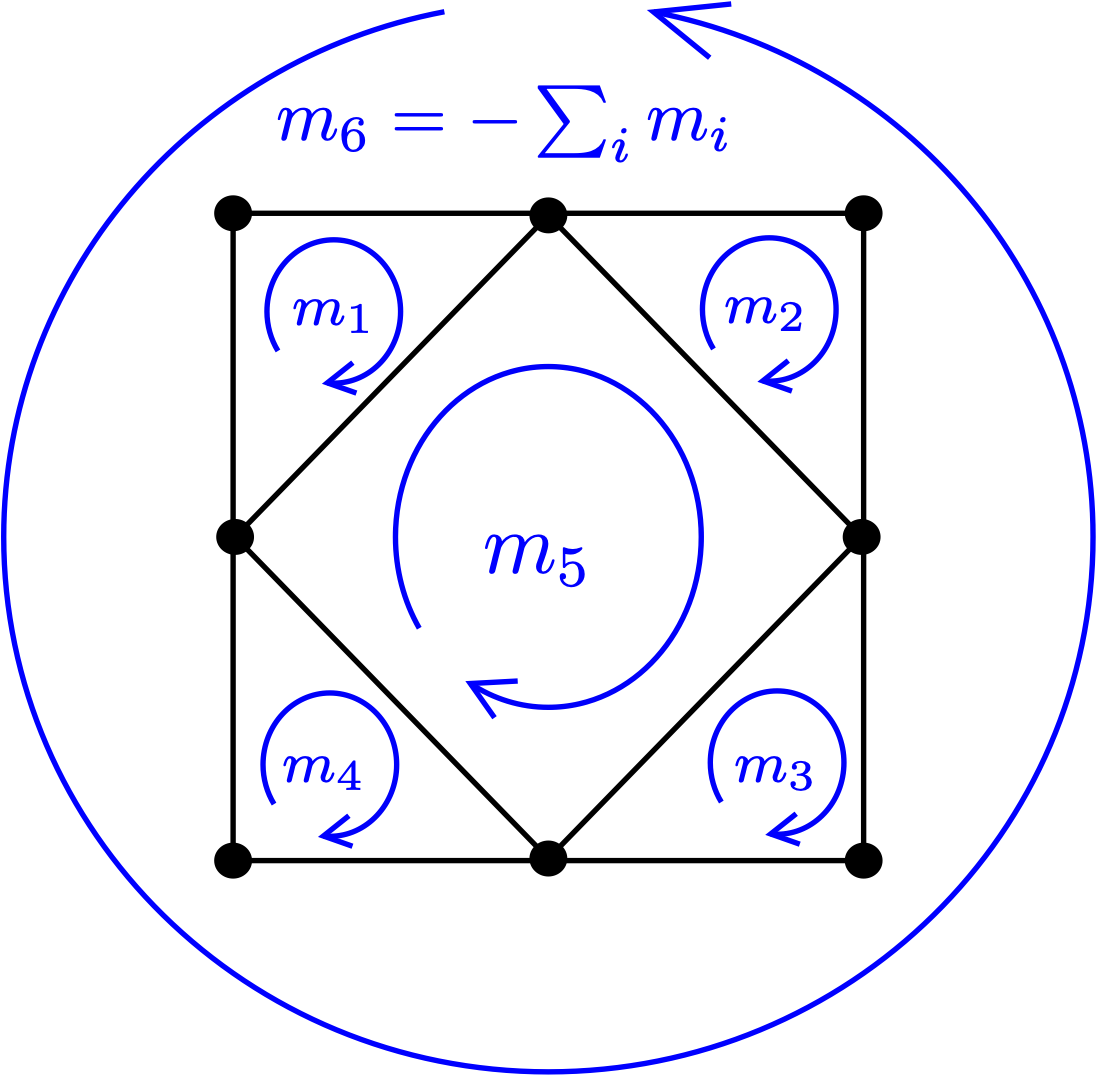}
		\end{minipage}
	\end{tabular}
\caption{
$\mathbb{Z}_M$ fluxes on Feynman diagrams: The left figure shows twists on links (i.e. propagators) $\{n_i\}$ and the flux 
on the plaquettte, which is given by 
a sum of twists around the plaquette $m=\sum_i n_i$ and  invariant under “local” $\mathbb{Z}_M$ gauge transformations on vertices.
A configuration of $\mathbb{Z}_M$-invariant  twists is given by the fluxes on plqeuttes (right).
%The right figure is an example of a set of $\mathbb{Z}_M$-invariant fluxes of twists on plaquettes.
}
\label{Fig1}
\end{figure}
%%%%%%%%%%%%%%%%%%%%%%%%%%%%%%%%%%%%%%%%%%%

On a $\mathbb{Z}_M$ orbifold, each propagator in a Feynman diagram is twisted as in Eq.(\ref{eq:green}) and it is oriented by the twist number $m\in\mathbb{Z} \mod M$. 
	%=-\lfloor \frac{M-1}{2} \rfloor, \cdots, \lceil \frac{M-1}{2} \rceil$, where $\lfloor x \rfloor \equiv \max \{n\in\mathbb{Z}\vert n\le x\}$ and $\lceil x \rceil \equiv \min\{n\in\mathbb{Z}\vert n\ge x\}$
%. 
As an example, see Fig.\ref{Fig1}. Four propagators with twist numbers $n_1,\cdots, n_4$ form one single loop. When we rotate the coordinates at a vertex $x$ by $2\pi l/M$, $n_1$ and $n_2$ are shifted by $l$ and $-l$, respectively.  
Therefore, 
the sum of twists around a plaquette $m=\sum_i n_i$, which we sometimes call a \textit{flux}, is invariant under $\mathbb{Z}_M$ rotations at vertices. %Consequently, 
It follows that for a given diagram such as the right figure of Fig.\ref{Fig1}, independent configurations of twists are characterized by twist numbers of plaquettes. This is why the number of independent twists coincides with the number of loops L. For convenience, we assign the complement twist number to the outer circle ($m_6=- \sum_{l=1}^5 m_l$ in the right figure of Fig.\ref{Fig1}), while it is not an independent twist. In the following, we omit writing such a twist assigned in an outer circle.

This prescription can be understood clearly when we interpret the twists as a kind of gauge fields. Regarding vertices in a Feynman diagram as topologically connected ``sites'' on a lattice.  
Since a twist on a propagator is defined between the two vertices, it can be seen as a link variable associated with the relative phase of the vertices. Then, the %invariance
$\mathbb{Z}_M$ rotation on each vertex is interpreted as a local change of the phase. It is nothing but a gauge transformation, 
but the angle  is restricted to ($2\pi m/M$) with $m=0,\cdots,M-1$. 
As a result, it is understood as {\it $\mathbb{Z}_M$ gauge theory on Feynman diagrams.} 
A flux in a plaquette, namely a sum of twists around the plaquette,  %which 
is invariant under $\mathbb{Z}_M$ rotations at vertices and characterizes  distinct configurations.
Thus it is a counterpart of the Wilson loop, a gauge-invariant object in a gauge theory. 
%A difference is that the internal space in the present case is phases of the coordinates or momenta. 
%Accordingly, a twist of a plaquette is also regarded as a twist of the corresponding loop momentum. 
A flux of twists is defined  
as a sum of the twists  of propagators in a counterclockwise direction along a plaquette. %and 
%It is equivalent to consider the twisted loop momentum flowing in the same dirrection. 
The flux is, of course, defined modulo $M$; i.e. $-m$ flux is equivalent to $M-m$ flux.

The procedure %to calculate a contribution to 
to calculate EE  is straightforward: 
perform momentum integrations of each bubble Feynman diagram 
with a fixed configuration of twists, i.e. fluxes $\{m_l\}$ for plaquettes, 
and sum up them over all the twist configurations  of $m$'s. 
Then sum all the bubble diagrams as usual to obtain the free energy. 
Since the configuration of trivial twists does not contribute to EE as discussed in the previous section, 
we are interested in a configuration of twists, in which some of them are non-vanishing. 
An evaluation of  Feynman diagrams with nonvanishing twists is in principle 
straightforward but very involved
since  momenta are twisted. 
Thus our strategy is,  instead of considering general configurations of twists, 
to focus on dominant contributions to EE.
In particular,  in the following sections, 
we consider two specific types of configurations, giving 
contributions from {\it twisted propagators} 
and those from {\it twisted vertices}. 
We discuss in Sec.\ref{s:discussion} why they will give dominant contributions to EE
and how  the rest of contributions are incorporated in the Wilsonian renormalization picture.\footnote{
	Although in Sec. \ref{s:prop} and \ref{s:vert}, flux configurations that cannot be attributed to a single twist of either a propagator or a vertex remain uncalculated, we can address this issue via the Wilsonian renormalization group, which will be discussed in Sec. \ref{s:discussion}.
}

%%%%%%%%%%%%%%%%%%%%%%%%%%%%%%%%%%%%%%%%%%%%%%%
\section{Propagator contributions to EE}
\label{s:prop}
\begin{figure}[t]
\centering
\includegraphics[width=9cm]{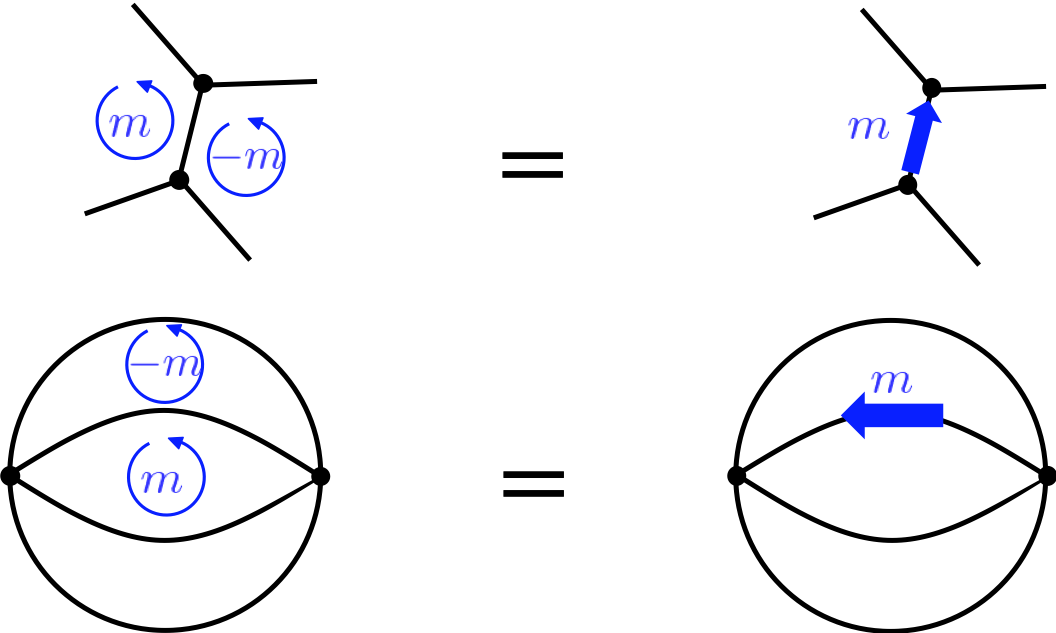}
\caption{
If fluxes of plaquettes straddling a shared propagator are given by $m$ and $-m$, 
such a configuration is interpreted  as a  twist of the shared %common
propagator.   
The upper figures show a relevant part with the  twisted propagator  in general diagrams.
%denote a part of such a diagram  are inserted in general diagrams.
}
\label{f:proptwist}
\end{figure}

%%%%%%%%%%%%%%%%%%%%%%%%%%%%%%%%%%%%%%%%%%%%%%%
Among various configurations of  twists, we first focus on the configurations that a single propagator is twisted. 
Consider a configuration where  two plaquettes with a nonvanishing flux of twists 
share a  propagator 
and their fluxes are given by $m$ and $-m$ respectively.
For such a diagram, both of the fluxes  can be attributed to the 
$m$-twist of the shared propagator  (Fig.\ref{f:proptwist}) and we can 
interpret such a flux configuration as a twist of the propagator.
%%%%%%%%%%%%%%%%%%%%%%%%%%%%%%%%%%%%%%%%
%%%%%%%%%%%%%%%%%%%%%%%%%%%%%%%%%%%%%%%%%%
The contributions to EE from such a class of diagrams are then understood as two-point function contributions. %, and 
We will investigate it both in the perturbative and nonperturbative approaches.

\begin{figure}[t]
\centering
\includegraphics[width=4cm]{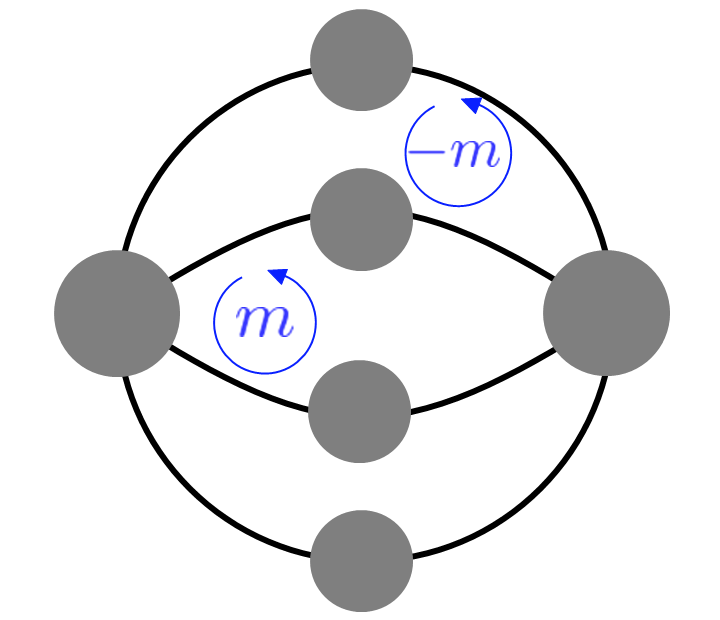}
\caption{
An example of a configuration of fluxes $(m,-m)$ that has multiple  interpretations 
in terms a twist of a {\it bare} propagator. 
The gray blobs %denote 
represent 1PI subdiagrams. 
This configuration can be interpreted as a twist of  either 
one of the two shared (bare) propagators, but not both. }
\label{f:ambiguity1}
\end{figure}

One might suspect whether  a configuration of fluxes like Fig.\ref{f:proptwist} has a one-to-one correspondence
to a configuration of  a twist of the propagator   in general diagrams. 
Indeed, we need careful treatment for particular diagrams. 
Consider a diagram like Fig.\ref{f:ambiguity1} 
where two  plaquettes with nonzero fluxes meet at two or more propagators. 
In this case,  the configuration of fluxes $(m,-m)$ 
corresponds to a twist of either propagator, but not to both. 
This example shows that such a configuration of fluxes can be interpreted as a twist of the  full propagator. 
In the following, we investigate propagator contributions to EE in more detail. 
%
%%%%%%%%%%%%%%%%%%%%%%%%%%%%%%%%%%%%%%%%
%%%%%%%%%%%%%%%%%%%%%%%%%%%%%%%%%%%%%%%%%%%

%%%%%%%%%%%%%%%%%%%%%%%%%%%%%%%%%%%%%%%%%%%
%%%%%%%%%%%%%%%%%%%%%%%%%%%%%%%%%%%%%%%%%%%
\subsection{Twisted propagator as a pinned propagator at the boundary}
\label{sec:twist}
\begin{figure}[t]
\centering
\includegraphics[width=8cm]{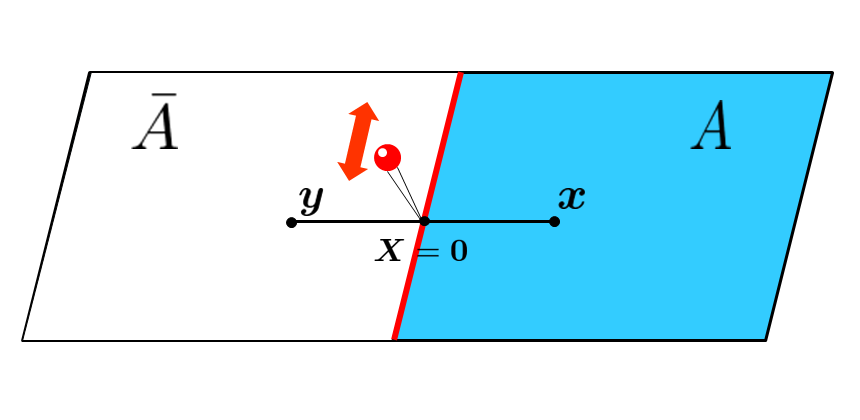}
\caption{
An illustrative picture of a propagator pinned on the boundary. Its midpoint $\bm{X}=(\bm{x}+\bm{y})/2$ is constrained on the boundary $\partial A$
while two end points move freely.
}
\label{f:pinnedprop}
\end{figure}

Before proceeding to an investigation of  individual diagrams, we %shall 
address a concrete interpretation of a twisted propagator
 in order to get a physical intuition for twisting.
We demonstrate below that a twisted propagator is  pinned at the boundary. 
For this purpose, 
it is convenient to introduce the center-of-mass and relative coordinates: $X=(x+y)/2$, $r=x-y$. 
A twisted propagator with $m\neq 0$ in the position space is written as
\aln{
G_{0} (\hat{g}^m x -y)  &=  
G_{0} ( \hat{g}^{m/2} \bm{x} - \hat{g}^{-m/2} \bm{y} ; r_\parallel) % \nonumber \\ 
= G_0 (\cos \theta_m \bm{r} + 2 \sin \theta_m (\epsilon \bm{X}) ; r_\parallel) 
\nonumber \\
&= e^{\cot \theta_m \hat{R}_{\bm{X}} /2} G_0 (2 \sin \theta_m  \bm{X} ;  r_\parallel) \nonumber\\
&=\frac{e^{\cot \theta_m \hat{R}_{\bm{X}} /2} }{4 \sin^2 \theta_m}  
  \int \frac{ d^{d-1} k_\parallel}{(2\pi)^{d-1}}
	\frac{	e^{ i k_\parallel \cdot r_\parallel}}{\left(-\partial^2_{\bm{X}}/4 \sin^2 \theta_m \right) + M^2_{k_\parallel}} 
\delta^2 (\bm{X}) ,
\label{twistedpropagator}
}
where 
\aln{
[\epsilon \bm{X}]_i &= \sum_j \epsilon_{ij}X_j \qq{($\epsilon_{ij}$: {the two-dimensional Levi-Civita symbol}),} \\
\hat{R}_{\bm{X}}&=\bm{r}\cdot(\epsilon \partial_{\bm{X}}), \  \
\theta_m= \frac{m\pi}{M},  \  \
M_{k_\parallel}^2 =k_\parallel^2+m^2_{0}.
}
Eq.\eqref{twistedpropagator} can be written in  a derivative expansion  on the delta function
with respect to $\partial^2_{\bm{X}}/M^2_{k_\parallel}$. 
When we consider a diagram with a single twist $m$ on a  %either of the 
propagator, it is formally written as
\aln{
\int d^{d+1}x\, d^{d+1}y\, G_0(\hat{g}^mx-y)F(r),\qq{where} r=x-y.
}
The integrand other than the twisted propagator only depends on $r$ due to the translational invariance.
%It is due to the translation invariance in the part other than the twisted propagator. 
With Eq.(\ref{twistedpropagator}) and the partial integration, we can drop all the $\partial_{\bm{X}}$ in the expression. 
Therefore, in this case, we can replace the propagator in the diagram as
\aln{
G_0(\hat{g}^mx-y)&\rightarrow\frac{1}{4\sin^2\theta_m}\int\frac{d^{d-1}k_\parallel}{(2\pi)^{d-1}}
\frac{e^{ik_\parallel\cdot r_\parallel}}{M_{k_\parallel}^2}
\delta^2(\bm{X})\nonumber\\
&=\frac{1}{4\sin^2\theta_m}G_0^{\text{bdry}}(r_\parallel)\delta^2(\bm{X}).
\label{e:efftwistprop}
}
%Here, 
$G_0^{\text{bdry}}$ is an ordinary propagator but 
its propagation is restricted only to the directions parallel to the boundary. 
Now the physical meaning of Eq.(\ref{e:efftwistprop}) is clear. Since the boundary of the subregion rests at the origin of the orbifold, %the twisted propagator has its 
the midpoint $\bm{X}$ of the twisted propagator is constrained on the boundary. Note that the propagator itself is not trapped on the boundary since the relative coordinate ${\bm r}$ is not constrained at all.   Rather, ${\bm r}$-dependence 
completely disappears from the twisted propagator.
Hence, it can be seen as a ``pinned'' propagator with  two {\it loose ends} on the two-dimensional plane.
This shows that the twisted propagator reflects a 
correlation between two points that are symmetrically distant from the boundary (Fig.\ref{f:pinnedprop}). 
In this sense, we can identify contributions to EE from a single twisted propagator as the
quantum correlation of two-point functions.  
In Eq.(\ref{eq:twist-k}), we saw that the one-loop contribution to EE from a twisted propagator 
gives a nontrivial delta function, $\delta^2 ({\bm k})$, which is responsible for the area law of EE.
Thus the above observation gives a different interpretation for the area law. 
%%%%%%%%%%%%%%%%%%%%%%%%%%%%%%%%%%%%%%%%
%%%%%%%%%%%%%%%%%%%%%%%%%%%%%%%%%%%%%%%%%%%

\medskip

%%%%%%%%%%%%%%%%%%%%%%%%%%%%%%%%%%%%%%%%%%%
%%%%%%%%%%%%%%%%%%%%%%%%%%%%%%%%%%%%%%%%%%%
\subsection{Perturbative analysis}
\begin{figure}[t]
\centering
\includegraphics[width=0.25\linewidth]{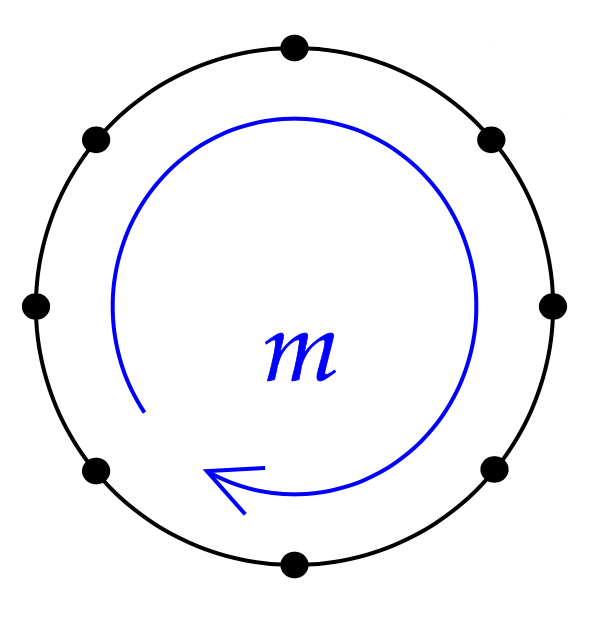}
\caption{
There is a single twist for the 1-loop diagram consisting of a product of propagators.}
\label{Fig2}
\end{figure}

We now investigate various diagrams containing a twisted propagator.
Let us begin with 1-loop diagrams as shown in Fig.\ref{Fig2}. 
In the perturbative approach, they are given by the Feynman diagrams with {“two-point vertices,”}
\aln{
		\Tr \log [\hat{P}G_0^{-1}\hat{P}]=\int_{{\epsilon^2}}\frac{ds}{s}\Tr e^{-sG_0^{(M)}/M}=\int_{\epsilon^2}\frac{ds}{s}\sum_{n=0}^{\infty}\frac{(-s)^n}{n!}\Tr\left[\left(G_0^{(M)}\frac{1}{M}\right)^n\right].
\label{e:1loopexpand}
} 
%They are exceptional 
These diagrams are exceptional in the sense that they are composed of %consist of 
a {single} chain of the propagators {connected by the two-point vertices of $(1/M)$}.\footnote{%-----------------------
	{Reflecting the orbifold action in (\ref{eq:action}), the path integral measure is given by 
		\aln{
			\int \mathcal{D}(\delta\phi)\, e^{-\frac{1}{2M}\int d^{d+1}x(\delta\phi)^2}=1
		} 
		so that the 
		1-loop part of the free energy
		is given by $(1/2)\mathrm{Tr}\,\mathrm{ln}( \hat{P} G_0^{-1} \hat{P}) = (1/2)\mathrm{Tr}\,\mathrm{ln}(M(G_0^{(M)})^{-1})$
		rather than $(1/2)\mathrm{Tr}\,\mathrm{ln}((G_0^{(M)})^{-1})$. 
		See Eq.(\ref{propagator-orbifold}). 
		This is responsible for the coefficient $(1/M)$ of the “two-point vertex”  in Eq.(\ref{e:1loopexpand}). 
	}
} %--------------------
%(Fig.\ref{Fig2}). 
There is only one configuration with $m$-flux  for the center plaquette. 
It can be regarded as a twist of a {\it single} propagator among
$n$ propagators in the expansion Eq.(\ref{e:1loopexpand}).
 %\footnote{%-------------------
%Note that they are in the same kind of configurations as Fig.\ref{f:proptwist} with the complementary twist assigned to the outer circle. 
%} %--------------------- 
This is what we have mentioned at the beginning of this section. %above. 
From the viewpoint of operators, it corresponds to the idempotency of the projection: $\hat{P}^2=\hat{P}$. 
%Therefore, $G_0^{(M)}$'s in Eq.(\ref{e:1loopexpand}) can  be replaced with an untwisted propagator $G_0$ 
%except one at each diagram. 

%%%%%%%%%%%%%%%%%%%%%%%%%%%%%%%%%%%%%%%%
%%%%%%%%%%%%%%%%%%%%%%%%%%%%%%%%%%%%%%%%%%%
%Since twists do not affect transverse $(d-1)$-dimensional space, we write only the 2-dimensional 
%momenta (or coordinates) in the following, unless otherwise noted. 
%The free energy with twist $n$ is easily calculated \cite{Nishioka_2007} by noting 
%that $\bra{\bm{k}} \hat{g}^n \ket{\bm{k}}= (2\pi)^2 \delta^2 (\bm{k}) /4 \sin^2(n \pi /M)$ for $n \neq 0$. 

In this case, it is  convenient to take the momentum space representation of a twisted propagator, 
instead of the above position space interpretation.
 There is a single loop momentum $k$ and  twisting results in a nonvanishing delta function of 
  $\delta^2(\hat{g}^m\bm{k}-\bm{k})$. The free energy is calculated as
\aln{
\tilde{F}^{(M)}_\text{1-loop}
&=
\frac{1}{2}\mathrm{Tr}\,\mathrm{ln}\,(\hat{P}G_0^{-1}\hat{P})
%\nonumber\\ &
=\frac{1}{2M}
\int_{\epsilon^2}\frac{ds}{s}\sum_{n=0}^{\infty}\frac{(-s)^n}{n!}\mathrm{Tr}\,[G_0^{n-1}\,G_0^{(M)}]\nonumber\\
&=\frac{1}{2M}
\sum_{m=0}^{M-1}\int_{\epsilon^2}\frac{ds}{s}\sum_{n=0}^{\infty}\frac{(-s)^n}{n!}\int\frac{d^2\bm{k}}{(2\pi)^2}\frac{d^{d-1}k_\parallel}{(2\pi)^{d-1}}\left(\frac{1}{k^2+m^2_{0}}\right)^n
(2\pi)^{d+1}
\delta^2(\hat{g}^m\bm{k}-\bm{k})
\delta^{d-1}(0)\nonumber\\
&= \frac{1 }{2 M} \int \frac{d^2 \bm{k} \ d^{d-1}k_{\parallel}} {(2 \pi)^{d-1}}   \log (k ^2 + m^2_{0}) \left( \frac{V_{d+1}}{(2\pi)^2}+  V_{d-1}\frac{M^2-1}{12}  \delta^2(\bm{k}) \right) 
\label{e:freeenergy1loop}
}
and Eq.(\ref{eq:1-loop-F}) is reproduced. 
Hence EE in the free theory is given by Eq.(\ref{EE-1loop}). 
In terms of the restricted propagator $G_0^{\text{bdry}}$ on the boundary, it is written as
\aln{
S_{\text{1-loop}}=-\frac{V_{d-1}}{12}\int^{1/\epsilon}\frac{d^{d-1}k_\parallel}{(2\pi)^{d-1}}\mathrm{ln}\,\left[\bigl(\tilde{G}_0^\text{bdry}(k_\parallel)\bigr)^{-1}\epsilon^{2}\right],
\label{e:S1loop}
} 
where $\tilde{G}_0^{\text{bdry}}$ is the momentum space representation of $G_0^{\text{bdry}}$.
$\bigr(\tilde{G}_0^\text{bdry}(k_\parallel)\bigr)^{-1}=k_\parallel^2+m^2_{0}$ is  %a  %mass squared of 
%the nonzero $k_\parallel$ momentum field on the two-dimensional plane.
{an effective squared mass on the two-dimensional plane with nonzero transverse momentum $k_\parallel$.}

\medskip

Next, we study contributions to EE from multi-loops. 
Flux configurations of the 2-loop figure-eight diagram are characterized by
 twists $(m_1, m_2)$ on the two plaquettes. 
Its contribution to the free energy is given by 
\aln{
\tilde{F}_{\text{2-loop}}^{(M)}= 
\sum_{m_1,m_2} \frac{3 \lambda}{4M} \int d^{d+1} x \ G_0 (\hat{g}^{m_1} x, x) G_0(\hat{g}^{ m_2} x, x) .
\label{8figure}
}
%%%%%%%%%%%%%%%%%%%%%%%%%%%%%%%%%%%%%%%%
\begin{figure}[t]
\centering
\includegraphics[width=9cm]{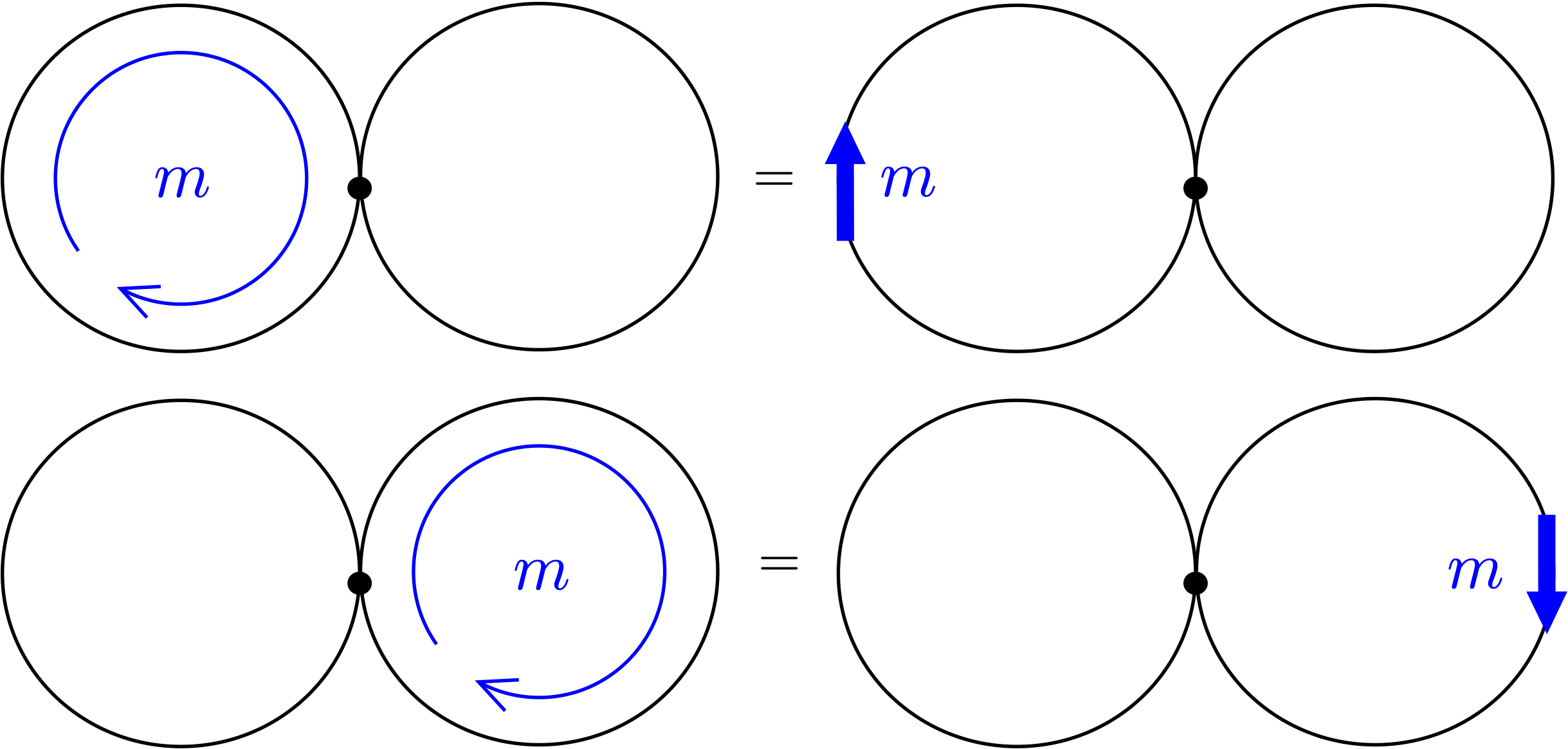}
\caption{
 2-loop diagram with twists $(m_1, m_2)=(m,0), (0,m)$ with $m\neq0$ (left). 
 They are interpreted as a twist of the corresponding propagators (right).
}
\label{Fig3}
\end{figure}
%%%%%%%%%%%%%%%%%%%%%%%%%%%%%%%%%%%%%%%%%%%
Specific configurations of twists, $(m, 0)$ and $(0, m)$ with $m\neq 0$, correspond to a twist of each propagator (Fig.\ref{Fig3}). By using Eqs.(\ref{e:efftwistprop}) and (\ref{e:summ}), 
their contributions to the free energy and EE are computed respectively as
\aln{
\tilde{F}_\text{2-loop, prop}^{(M)}&=2\times\frac{3\lambda}{4M}\sum_{m=1}^{M-1}\int d^{d+1}x\frac{1}{4\sin^2\theta_m}G_0^{\text{bdry}}(0)\delta^2(\bm{x})G_0(0)\nonumber\\
&=V_{d-1}\frac{3\lambda(M^2-1)}{24M}G_0(0)G_0^{\text{bdry}}(0),\\
S_\text{2-loop, prop}&=-\frac{V_{d-1}}{12}G^{\text{bdry}}_0(0)\,[3\lambda G_0(0)]\nonumber\\
&=
-\frac{V_{d-1}}{12}\int\frac{d^{d-1}k_\parallel}{(2\pi)^{d-1}}\tilde{G}_0^{\text{bdry}}(k_\parallel)[3\lambda G_0(0)]. 
\label{e:S2loopprop}
}
Note that the vertex contribution to EE is negative for the repulsive (positive $\lambda$) interaction. 
It is consistent with an expectation that the degrees of freedom must be reduced by introducing a positive $\lambda$
(otherwise the system becomes unstable) interaction.

%Here, 
Eq.(\ref{e:S2loopprop}) indicates that this contribution to EE can be attributed to the  mass renormalization 
to the 1-loop contribution of Eq.(\ref{e:S1loop}):
\begin{gather}
S_\text{1-loop}+S_\text{2-loop, prop}=
-\frac{V_{d-1}}{12}\int^{1/\epsilon}\frac{d^{d-1}k_\parallel}{(2\pi)^{d-1}}\mathrm{ln}\,\left[\bigl(\tilde{G}_1^{\text{bdry}}(k_\parallel)\bigr)^{-1}\epsilon^{2}\right],
\\
\tilde{G}_1^{\text{bdry}}(k_\parallel)=\frac{1}{k_\parallel^2+m_{r1}^2},~~~~
m_{r1}^2=m^2_{0}+ 3\lambda G_0(0).
\label{e:renormalizem}
\end{gather}
The above equalities hold up to $O(\lambda^1)$. This was also suggested in~\cite{Hertzberg:2012mn} to $O(\lambda^1)$.

%However, 
When we compute higher-order contributions by explicit calculations, we %get an observation
observe that the propagator contributions are absorbed in the ordinary renormalization of the propagator order by order. 
This fact comes from the property explained in Fig.\ref{f:ambiguity1} that $(m,-m)$ type configurations of fluxes
straddling  many consecutive bare propagators will twist the corresponding single {\it full} propagator. 
It is not a trivial fact, but physically  %but
natural since EE is a measure of entanglement among microscopic degrees of freedom and should be related to the low-energy observables through renormalization. This observation motivates us to pursue the following analysis
 that EE (or at least its universal term) is expressed in terms of renormalized correlation functions in the 2PI formalism.

%%%%%%%%%%%%%%%%%%%%%%%%%%%%%%%%%%%%%%%%%%
\subsection{Nonperturbative analysis in 2PI formalism}
In order to study a relationship between renormalization of propagators and EE more systematically, 
we employ the framework of the two-particle irreducible (2PI) formalism~\cite{PhysRevD.10.2428,Berges:2004yj}. 
Combined with the orbifold analysis, %As a result, 
we will confirm that the Gaussian contributions to EE are completely expressed in terms of the renormalized two-point function 
in the following.

The 2PI effective action is given by
\aln{
\Gamma[G]  =  \frac{1}{2} \tr \log G^{-1} + \frac{1}{2} \tr(G_0^{-1} G-1) +\Gamma_2[G],
\label{2PIEA}
}
where $G$ is a full propagator, namely, a renormalized two-point function. $\Gamma_2[G]$ is minus the sum of connected 2PI bubble diagrams which consist of the full propagators $G$'s as internal lines. We assume that the one-point function vanishes: $\left<\phi\right>=0$. In this formalism, $G$ is determined self-consistently by its equation of motion,  called a gap equation: \aln{
\frac{\delta \Gamma[G]}{\delta G}=0~~~\Leftrightarrow~~~G^{-1}=G_0^{-1}+2\frac{\delta \Gamma_2}{\delta G}[G].
\label{e:gapeq} 
} 
With the solution to Eq.(\ref{e:gapeq}), $G=\bar{G}[G_0]$, $\Gamma[\bar{G}]$ coincides with the 1PI free energy. 
Thus, what we need to evaluate is $\Gamma[\bar{G}]$ %consisting of diagrams 
with a single %twisted
full propagator being twisted. 

\vspace{5mm}
%%%%%%%%%%%%%%%%%%%%%%%%%%%%%%%%%%%%%%%%%%%%%%
%%%%%%%%%%%%%%%%%%%%%%%%%%%%%%%%%%%%%%%%%%%%%%

In the 2PI analysis, 
since  $G(x,y)$ itself is composed of  propagators as  internal loop corrections, 
we  distinguish the following two types of twistings.
The first type of twistings is denoted by $\delta_m G(x,y)$, which
represents a variation of the internal structure induced by twisting. 
The second type is simply given by $G(\hat{g}^mx,y)$, which represents the
twisting of the full propagator in the same way as previously. 
Namely, the projection operator $\hat{P}$ is acted from outside. 
We will show that the first type of twistings is canceled by the gap equation.
Moreover, we will prove that 
there are further cancellations among 2PI diagrams and the second term of the 2PI effective action in Eq.(\ref{2PIEA}). 
The gap equation is responsible for  the cancellations, but special care is necessary 
for such diagrams in Fig.\ref{f:ambiguity1}.

First let us see that twistings inside the full propagators  are canceled and
contributions from $\delta_m G(x,y)$ vanish. 
It is simply because of the gap equation;
\aln{
\Gamma[\bar{G}]_\text{prop,int}&=\sum_{m=1}^{M-1}\int d^{d+1}x\,d^{d+1}y\,\frac{1}{2}\left(-\bar{G}^{-1}+G_0^{-1}+2\frac{\delta\Gamma_2}{\delta G}[\bar{G}]\right)_{yx}\delta_mG(x,y)\nonumber\\
&=0.
}
Thus we can safely forget about the internal structure of the full propagator. 

Next, we look at the twisting of the full propagator itself.  
As expected, most configurations with a single twisted propagator
 are canceled due to the gap equation, except for diagrams like Fig.\ref{f:ambiguity1}
where a configuration of fluxes of $(m,-m)$ can be attributed to twisting  one of the propagators
straddled by the plaquettes. 
In the 2PI formalism,  such diagrams  are included only in the first term in Eq.(\ref{2PIEA})
 because all  diagrams with such property are not 2PI (see Fig.\ref{f:ambiguity1}) and not included in other terms.\footnote{
The second term is not 2PI, but $G_0^{-1}$ is a local operator and it is sufficient to twist the propagator $G$ in the trace. }
 Then, we can separately consider contributions from the first term 
 and those from the second and third term in Eq.(\ref{2PIEA}). 
 
The first term gives the same form of EE as in the free theory. 
A flux of twists is present in the center plaquette, 
 which can be attributed to one of the propagators, but not to all. 
 The situation is completely the same as in the 1-loop analysis in the previous section, and
  it results in the following contributions to EE,
\aln{
S^{\text{2PI}}_{\text{prop, ext,1}} =- \frac{V_{d-1} }{12} 
 \int^{1/\epsilon} \frac{d^{d-1}k_{\parallel}} {(2 \pi)^{d-1}}  
  \log \left[ \tilde{G}^{-1}(\bm{0}; k_\parallel) \epsilon^2 \right],
\label{EE-1loop2PI,1}
}
where $\tilde{G}(\bm{k}; k_\parallel)$ is the Fourier transform of $\bar{G}(x,y)$. 
$\tilde{G}(\bm{0}; k_\parallel)$ is a renormalized counterpart of $\tilde{G}^{\text{bdry}}$. 
Note that,  though
$\tilde{G}_0^{\text{bdry}}(k_\parallel)$ describes a propagation in a $(d-1)$-dimensional theory, 
the renormalization of $\tilde{G}(\bm{0}; k_\parallel)$ itself is performed in 
the $(d+1)$-dimensional space, as in Eq.(\ref{e:renormalizem}).

As for the second and third terms in Eq.(\ref{2PIEA}), %they give a contribution 
their contributions to EE are given by
\aln{
S^{\text{2PI}}_\text{prop,ext,2+3}&=\sum_{m=1}^{M-1} 
\int d^{d+1}x \,d^{d+1}y\,
\left (\frac{1}{2} G_0^{-1} +  \frac{\delta \Gamma_2}{\delta G}[\bar{G}] \right)_{yx} \bar{G}(\hat{g}^{m}x, y)\nonumber\\
&= \sum_{m=1}^{M-1} 
\int d^{d+1}x d^{d+1}y
\left (\frac{1}{2} \bar{G}^{-1} \right)_{yx} \bar{G}(\hat{g}^{m}x, y)%G(\hat{g}^{n}x, y).
\label{trivialEE}
}
Since the last expression is just a variation of unity, $S^{\text{2PI}}_\text{prop,ext,2+3}$ is a trivial constant and %we can drop it
can be dropped.

%%%%%%%%%%%%%%%%%%%%%%%%%%%%%%%%%%%%%%%%%%%%%%
%%%%%%%%%%%%%%%%%%%%%%%%%%%%%%%%%%%%%%%%%%%%%%

By combining %the above results 
Eqs.\eqref{EE-1loop2PI,1}, \eqref{trivialEE}, we obtain the contribution to EE from twisting a propagator in terms of %from 
the renormalized two-point function nonperturbatively:
\aln{
S^{\text{2PI}}_{\text{prop}} =- \frac{V_{d-1} }{12} 
 \int^{1/\epsilon} \frac{d^{d-1}k_{\parallel}} {(2 \pi)^{d-1}}  
  \log \left[ \tilde{G}^{-1}(\bm{0}; k_\parallel) \epsilon^2 \right]. 
\label{e:EE2PIprop}
}
%This completes a proof of the 
Previously we made a conjecture that the total propagator contribution to EE could be represented as renormalization of the propagator. 
The above argument completes the proof. %Note that 
The Gaussian contribution %from the Gaussian nature 
is all %organized 
summarized in the above form.\footnote{%-------------------------------------------
When we  compare Eq.(\ref{e:EE2PIprop}) to the ordinary perturbative calculation, 
since all the diagrams in Eq.(\ref{2PIEA}) are written in terms of the full propagator $G$, 
we have to expand  each diagram in the comparison. 
As a result, diagrams consisting of $G_0$'s are included in all the three terms in Eq.(\ref{2PIEA}) 
and the correct coefficients can be obtained by taking all these terms into account.
\label{fn:pertnonpert}
 } %-------------------------------- 
Note that it is consistent with the leading order result of perturbative calculations in \cite{Hertzberg:2012mn,Chen:2020ild}.

Though we have shown the above result for
the Gaussian contribution to EE in a specific model of the $\phi^4$ theory, 
a generalization to other theories is almost straightforward and  %it is a general result
the above result is completely general. 
See Section \ref{s:spin} for further details.

%%%%%%%%%%%%%%%%%%%%%%%%%%%%%%%%%%%%%%%%
\section{Vertex contributions to EE}
\label{s:vert}
%%%%%%%%%%%%%%%%%%%%%%%%%%%%%%%%%%%%%%%%
In the previous sections, we have successfully extracted the %complete 
Gaussian part of EE completely. The rest contributions to EE are purely non-Gaussian. 
Investigations of non-Gaussian contributions to EE are more involved 
since these contributions are hidden in various configurations of twists. 
However, some of them do have a simple interpretation as we show in this section.    

\subsection{Perturbative analysis}

\begin{figure}[t]
	\centering
	\includegraphics[width=10cm]{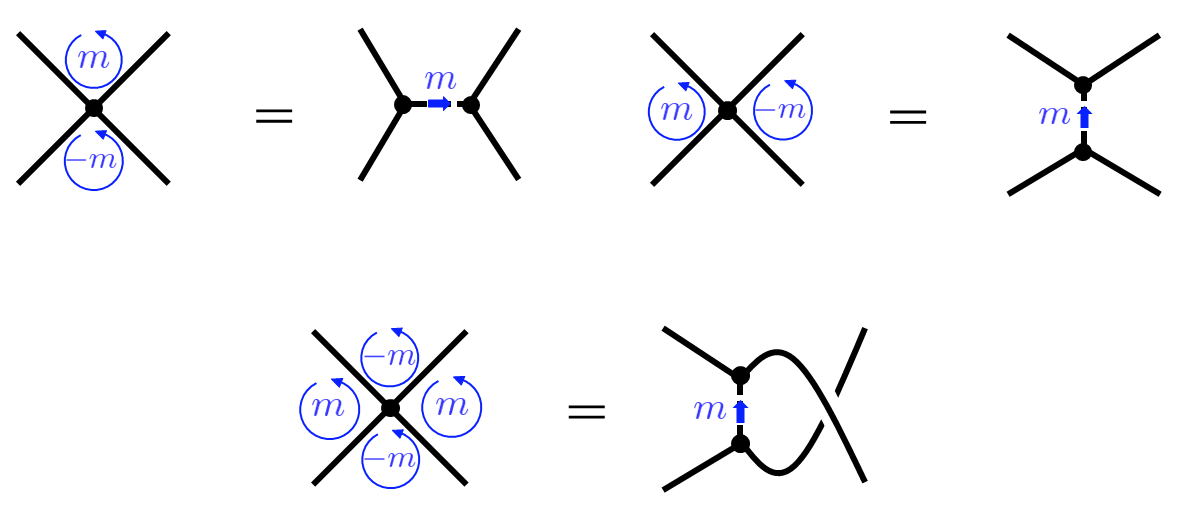}
	\caption{
Twisting a vertex: these three types of 
configurations can be attributed to a twist of a vertex. The dotted lines in the figures on the right-hand sides are delta functions to open the vertex. 
The twist of a vertex is interpreted as a twist of the dotted propagator. 
Each set of figures represent the three channels, 
$s$-channel (upper left figures), $t$-channel (upper right figures) and
$u$-channel (lower figures) respectively.
	}
	\label{f:verttwist}
\end{figure}

Such configurations with a simple interpretation are given by 
a set of flux configurations that {\it straddle a vertex} instead of a propagator. 
Consider a diagram with twists given schematically in Fig.\ref{f:verttwist}.
%%%%%%%%%%%%%%%%%%%%%%%%%%%%%%%%%%%%%%%%%%  
%%%%%%%%%%%%%%%%%%%%%%%%%%%%%%%%%%%%%%%%%%
%There, 
In these configurations, 
plaquettes with nonvanishing fluxes of twists 
meet at a vertex, and there are three types of such configurations. 
 We can interpret these configurations as a configuration of a single twisted vertex in the $s$, $t$, and $u$-channel respectively.
 This interpretation can be 
  realized by ``opening'' the vertex with a delta function. 
   For example, the four-point vertex can be rewritten as
\aln{
	\frac{\lambda}{4}\int d^{d+1}x\,\phi(x)^4=\frac{\lambda}{4}\int d^{d+1}x\,d^{d+1}y\,\phi(x)^2\phi(y)^2\delta^{d+1}(x-y).
}
Then,  we can understand a twisted vertex as %means 
an opened vertex with a twist on the separated two coordinates as
\aln{
	\frac{\lambda}{4}\int d^{d+1}x\,d^{d+1}y\,\phi(x)^2\phi(y)^2\delta^{d+1}(\hat{g}^mx-y).
} 
%The upper and lower figures in Fig.\ref{f:verttwist} correspond to the $s$- and $u$-channel openings of the vertex, respectively. 
%The $t$-channel opening is given in the same manner
%as the $s$-channel. 
The upper left, upper right, and lower figures in Fig.\ref{f:verttwist} correspond to the $s$, $t$, and $u$-channel openings of the vertex, respectively. 
As we have 
demonstrated for a single twisted propagator, 
we can replace the twisted delta function (to be exact, its two-dimensional part) in the diagram as 
\aln{
	\delta^{2}(\hat{g}^{m}\bm{x} - \bm{y}) =
	e^{\cot \theta_n \hat{R}_{\bm{X}} /2} \frac{ \delta^2 (\bm{X})}{4 \sin^2\theta_m}  \rightarrow \frac{\delta^2 (\bm{X})}{4 \sin^2\theta_m}. 
}
The {\it twisted vertex} is thus 
interpreted as a vertex symmetrically splitted with two loose ends and also  with its center coordinate being fixed at the boundary.

\begin{figure}[t]
	\centering
	\includegraphics[width=0.8\linewidth]{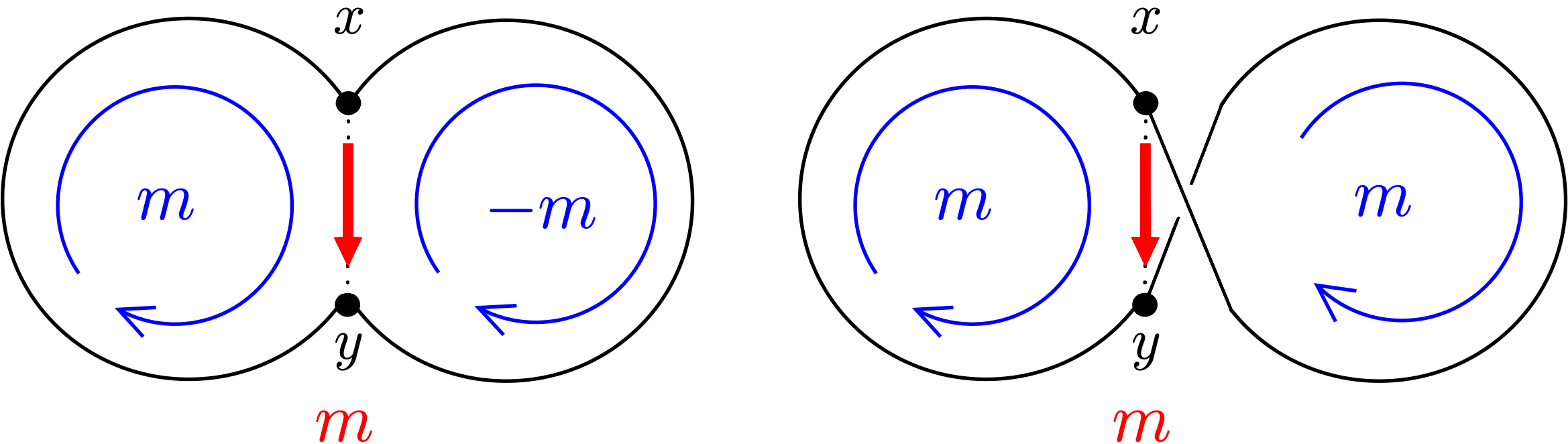}
	\caption{
		{
			2-loop figure-eight diagrams with twists $(m_1, m_2)=(m,\mp m)$. These flux configurations of  twists can be interpreted 
			as a twist of the 4-point vertex by decomposing it into two 3-point vertices. % by $\delta^{d+1}(x-y)$.
		}
	}
	\label{Fig3-2}
\end{figure}

Let us evaluate these vertex contributions up to the 3-loop level. The 2-loop vertex contributions stem
 from the figure-eight diagram with two types of configurations of twists, as shown in Fig.\ref{Fig3-2}. 
%%%%%%%%%%%%%%%%%%%%%%%%%%%%%%%%%%%%%%%%
%%%%%%%%%%%%%%%%%%%%%%%%%%%%%%%%%%%%%%%%%%%
Note that the configuration of the $s$-channel opening is absent in the figure-eight diagram
because the vertex in the figure-eight diagram is surrounded by essentially three plaquettes, %: 
two circles, and one outer circle.
% and in the $s$-channel diagram two pairs of plaquettes with simultaneous twists adjacent to each other, necessary for a vertex twist, is absent.  
Their contributions to the free energy and EE are calculated as
\aln{
	\tilde{F}^{(M)}_{\text{2-loop.vert}}&=2\times\sum_{m=1}^{M-1}\frac{3\lambda}{4M}\int d^{d+1}x\,d^{d+1}y\,G_0(x-y)^2\delta^{d-1}(x_\parallel-y_\parallel)\delta^2(\hat{g}^m\bm{x}-\bm{y})\nonumber\\
	&=V_{d-1}\lambda\frac{M^2-1}{8M}\int d^2\bm{r}G_0(\bm{r},0)^2,\\
	S_{\text{2-loop, vert}}&=-V_{d-1}\frac{\lambda}{4}\int d^2\bm{r}G_0(\bm{r},0)^2.
	%\nonumber\\&=-V_{d-1}\frac{\lambda}{4}\int d^{d+1}r\,G_0(\bm{r},0)^2\delta^{d-1}(r_\parallel).
	\label{e:EE2loopvert}
}
%In the scond line of Eq.(\ref{e:EE2loopvert}), we have rewritten the contribution for later use. 
Note that in the real $\phi^4$ theory, %the dotted lines are indistinguishable between different channels
different channels are indistinguishable 
and a summation of different channels give just an additional  numerical factor in front. 
In the next section, we will consider an extended model in which a different channel gives a different 
type of contribution.

\begin{figure}[t]
	\includegraphics[width=8cm]{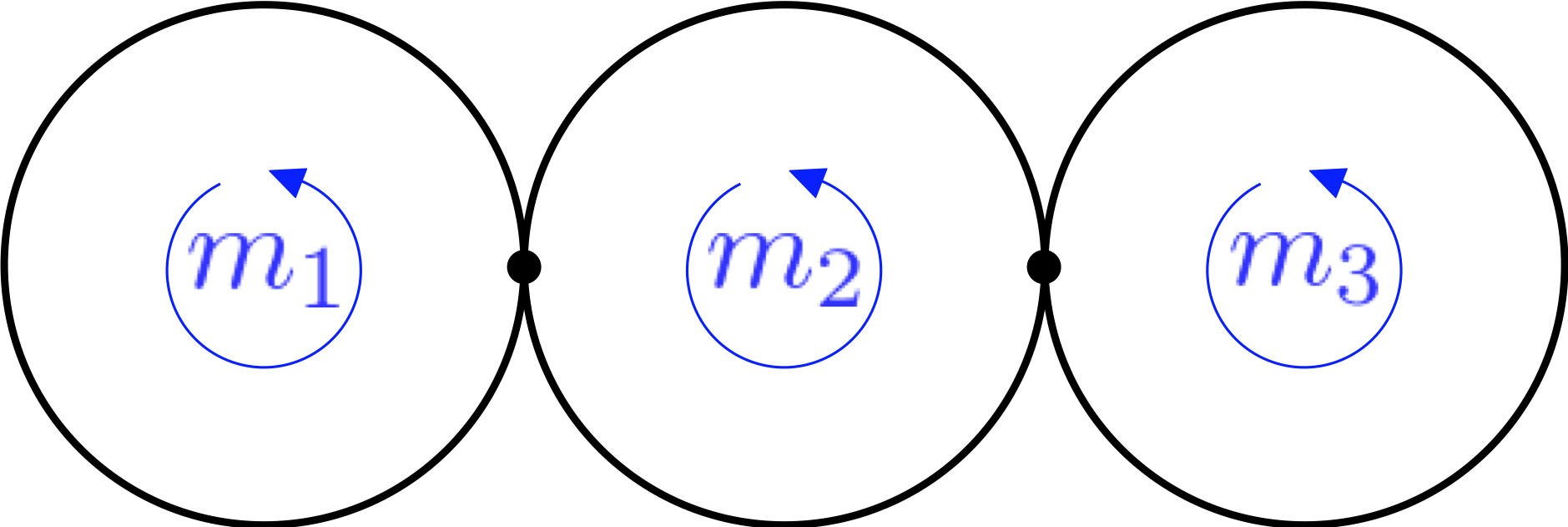}
	\caption{
		A 3-loop diagram. Four types of flux configurations,  $(m_1,m_2,m_3)=(m,\pm m,0)$, $(0,m,\pm m)$,
		can be interpreted as twisting vertices. 
		Opening vertices are done in the same manner as in the 2-loop diagrams.  
	}
	\label{f:3-loop-3}
\end{figure}

\begin{figure}[t]
	\begin{tabular}{c}%prevent line break
		\hspace*{-0.05\linewidth}
		\begin{minipage}{0.5\hsize}%align figs horizontally
			\centering
			\includegraphics[width=\linewidth]{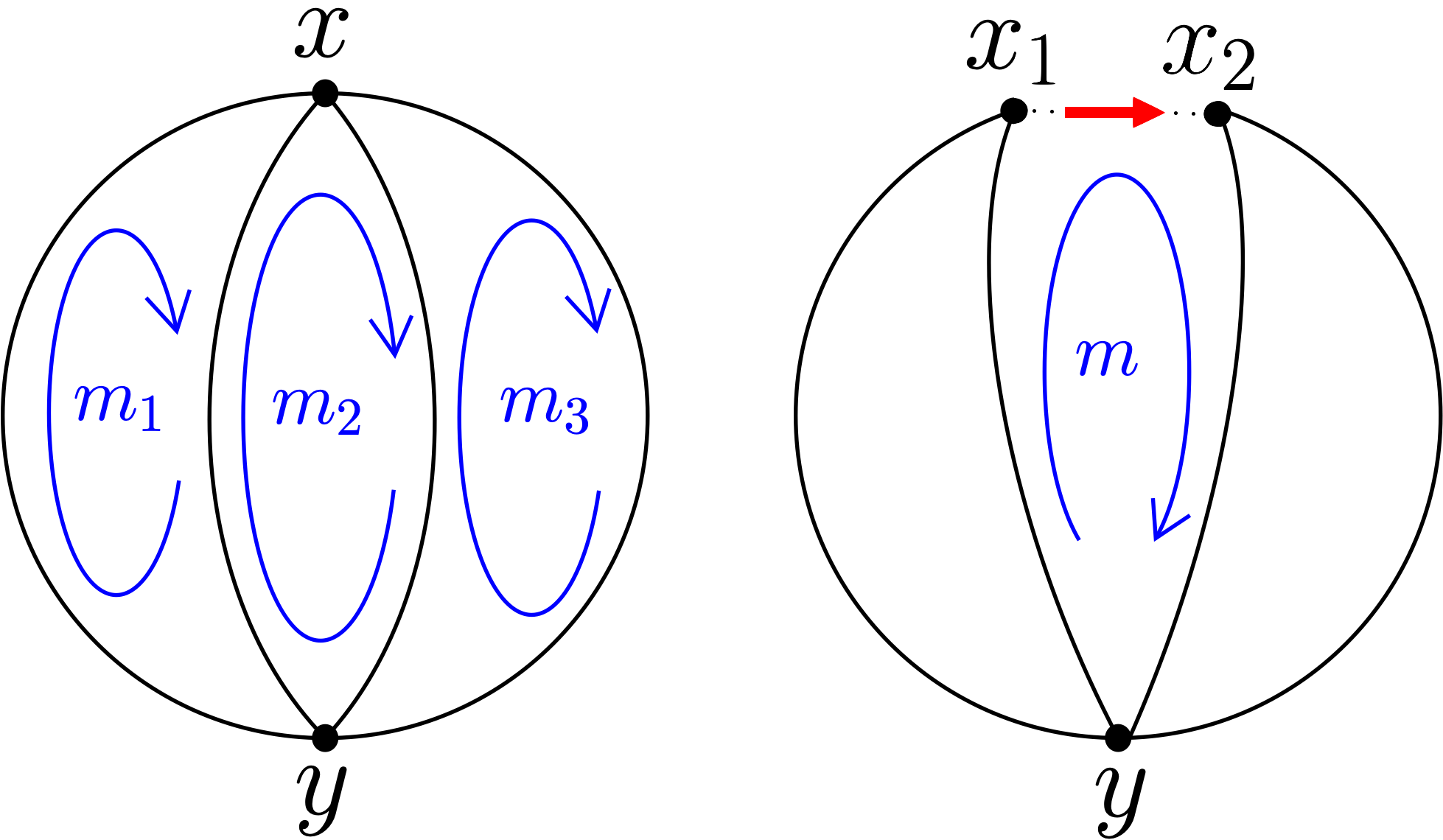}
		\end{minipage}
		\hspace*{0.03\linewidth}
		\begin{minipage}{0.5\hsize}
			\centering
			\includegraphics[width=\linewidth]{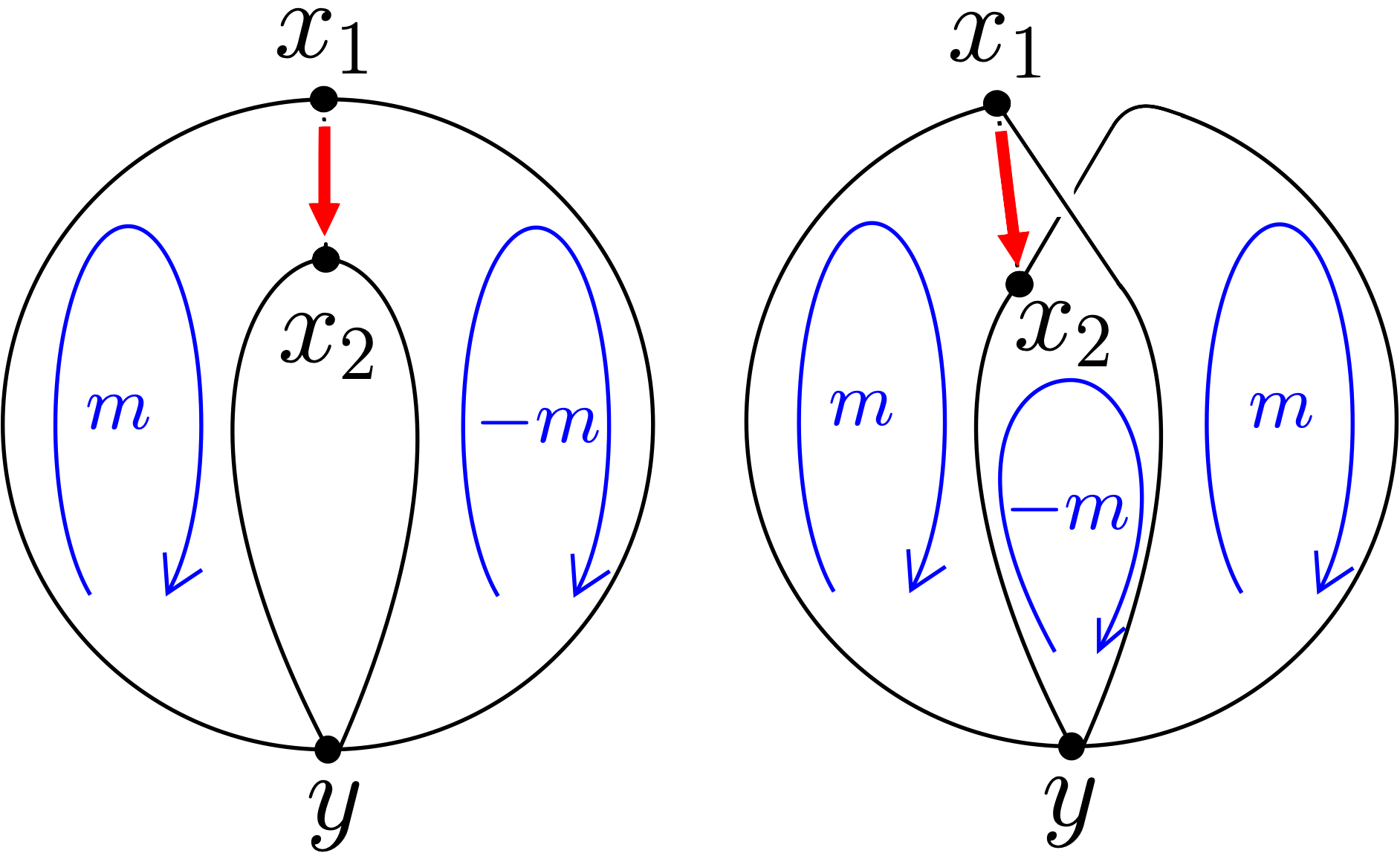}
		\end{minipage}
	\end{tabular}
	%\centering
	%\includegraphics[width=0.8\linewidth]{Fig5.png}
	\caption{
		Another 3-loop diagram with twists $(m_1, m_2, m_3)$ {(leftmost)}.
		A particular configuration $(0,m,0)$ corresponds to twisting a vertex, as well 
		as $(m,0,-m)$ and $(m,-m,m)$ (three diagrams on the right).  They generate a twist in the delta function  
		$\delta^2({\bm x}_1-{\bm x}_2)$. We can also open the vertex at ${\bf y}$ instead of ${\bm x}$, 
		and they have two different interpretations of twisting vertices, analogous to Fig.\ref{f:ambiguity1}.
		These vertex contributions should not be double-counted. 
		% of twists, $(m,0,\pm m)$ and $(0, m, 0)$,  corresponding to twisting a vertex (right).
	}
	\label{Fig5}
\end{figure}

\vspace{5mm}
The 3-loop contributions come from two diagrams shown in Fig.\ref{f:3-loop-3} and Fig.\ref{Fig5}. %One of them is illustrated in Fig.\ref{f:3-loop-3}. 
%%%%%%%%%%%%%%%%%%%%%%%%%%%%%%%%%%%%%%%%
%%%%%%%%%%%%%%%%%%%%%%%%%%%%%%%%%%%%%%%%%%%
%There, 
For a diagram illustrated in Fig.\ref{f:3-loop-3}, the vertex contributions stem from the four configurations: $(m_1,m_2,m_3)=(m,\pm m,0)$, $(0,m,\pm m)$. We see them as $t$- and $u$-channel opening of the two vertices. $s$-channels are absent because each vertex is surrounded by two plaquettes and one outer circle, not four independent ones. The contributions from these configurations to the free energy and EE are given by 
\aln{
		\tilde{F}^{(M)}_{\text{3-loop, vert1}}&=4\times\biggl(-\frac{9\lambda^2}{4M}\sum_{m=1}^{M-1}\int d^{d+1}x_1\,d^{d+1}x_2\,d^{d+1}y\,G_0(x_1-x_2)\,G_0(x_1-y)\nonumber\\*
		&\hspace{6cm}\times G_0(x_2-y)\,G_0(0)\,\delta^{d+1}(\hat{g}^mx_1-x_2)\biggr)\nonumber\\*
		&=-V_{d-1}\frac{3\lambda^2(M^2-1)}{M}\int d^2\bm{x}\,d^2\bm{y}\,d^{d-1}r_\parallel\ G_0(2\bm{x},0)\,G_0(\bm{x}-\bm{y},r_\parallel)\nonumber\\*
		&\hspace{8cm}\times G_0(\bm{x}+\bm{y},r_\parallel)\,G_0(\bm{0},0),\nonumber\\*
		&=-V_{d-1}\frac{3\lambda^2(M^2-1)}{4M}\int d^2\bm{r}\,d^2\bm{s}\,d^{d-1}r_\parallel\ G_0(\bm{r},0)\,G_0(\bm{r}-\bm{s},r_\parallel)G_0(\bm{s},r_\parallel)\,G_0(\bm{0},0),\\*
		S_{\text{3-loop, vert1}}&=V_{d-1}\,\frac{3}{2}\lambda^2\int d^2\bm{r}\,d^2\bm{s}\,d^{d-1}r_\parallel\ G_0(\bm{r},0)\,G_0(\bm{r}-\bm{s},r_\parallel)\,G_0(\bm{s},r_\parallel)\,G_0(\bm{0},0),\nonumber\\*
		&=V_{d-1}\,\frac{3}{2}\lambda^2\int d^{d+1}r\,G_0(r)\,G_0(0)\left[\int d^2\bm{s}\,G_0(\bm{s},0)\,G_0(\bm{s}-\bm{r},r_\parallel)\right].
	\label{e:EE3loopvert2}
}

Another 3-loop diagram is given by the leftmost diagram in Fig.\ref{Fig5}.
%%%%%%%%%%%%%%%%%%%%%%%%%%%%%%%%%%%%%%%%
%%%%%%%%%%%%%%%%%%%%%%%%%%%%%%%%%%%%%%%%%%%
The following three types of configurations of twists correspond to twists of a vertex:  $(0,m,0)$, $(m,0,-m)$, and $(m,-m,m)$. 
We can assign a flux of twist $-m$, $0$, and $-m$ on the outer circle of the plaquette respectively. 
They are equivalent to the $t$-, $s$- and $u$-channel opening of the vertex.   
In this diagram,  we again  face the problem of the failure of one-to-one correspondence in Fig.\ref{f:ambiguity1}. 
There are two ways to attribute the flux configurations to twisting either an upper or lower vertex. 
These two attributions are not independent and 
we can only twist one of them.  
These three channels give the same contributions in the $\phi^4$ theory. %and 
Then, the corresponding 3-loop contributions from Fig.\ref{Fig5} are computed as
\aln{
		\tilde{F}^{(M)}_{\text{3-loop, vert2}}&=3\times\biggl(-\frac{3\lambda^2}{4M}\sum_{m=1}^{M-1}\int d^{d+1}x_1\,d^{d+1}x_2\,d^{d+1}y\,G_0(x_1-y)^2G_0(x_2-y)^2
		%   \nonumber\\* &\hspace{8.4cm}\times
		\delta^{d+1}(\hat{g}^mx_1-x_2)\biggr)\nonumber\\* 
		&=-V_{d-1}\frac{3\lambda^2(M^2-1)}{4M}\int d^2\bm{x}\,d^2\bm{y}\,d^{d-1}		r_\parallel\,G_0(\bm{x}-\bm{y},r_\parallel)^2\,G_0(\bm{x}+\bm{y},r_\parallel)^2,\nonumber\\
		&=-V_{d-1}\frac{3\lambda^2(M^2-1)}{16M}\int d^2\bm{r}\,d^2\bm{s}\,d^{d-1}r_\parallel\,G_0(\bm{r},r_\parallel)^2\,G_0(\bm{s},r_\parallel)^2,\\
		S_{\text{3-loop, vert2}}&=V_{d-1}
		\frac{3\lambda^2}{8} d^2\bm{r}\,d^2\bm{s}\,d^{d-1}r_\parallel\,G_0(\bm{r},r_\parallel)^2\,G_0(\bm{s},r_\parallel)^2,\nonumber\\
		&=V_{d-1}
		\frac{3\lambda^2}{8}
		\int d^{d+1}rG_0(r)^2\left[
		\int d^2\bm{s}\,G_0(\bm{s},r_\parallel)^2\right].
	\label{e:EE3loopvert}
}
In contrast to the twisting of propagators, both of the contributions of Eqs.\eqref{e:EE3loopvert2} and \eqref{e:EE3loopvert} 
essentially originate from the non-Gaussianity of the state. 
We also emphasize the importance of the covariant viewpoint as $\mathbb{Z}_M$ gauge theory on Feynman diagrams.  
If we take a special gauge and assign twists on specific links (propagators), we could not find out vertex contributions since they are hidden in the configurations with multiple twisted links.

While Eq.(\ref{e:EE3loopvert2}) can be interpreted as a contribution from the figure-eight diagram with the renormalized propagator Eq.(\ref{e:renormalizem}), Eq.(\ref{e:EE3loopvert}) cannot be absorbed into the renormalization of the propagator nor the vertex. 
The situation is different from the propagator contributions, nonetheless, it is consistent with the ordinary renormalization structure in another viewpoint. 
In the following, we will show that the above vertex contributions can be %identified to 
summarized as those from renormalized composite operators.

%%%%%%%%%%%%%%%%%%%%%%%%%%%%%%%%%%%%%%%%%%%%%%%%%%%%%
\subsection{Vertex contributions as correlations of composite operators}
\label{s:aux}
%%%%%%%%%%%%%%%%%%%%%%%%%%%%%%%%%%%%%%%%%%%%%%%%%%%%%
In order to formulate the ``opening of a vertex'' more systematically, it is instructive to consider a model where opening each vertex leads to distinct $s$-, $t$- and $u$-channels. One of such models is described by two complex scalars, whose action is given by 
%written as follows:
\aln{
I=\int\frac{d^{d+1}x}{M}\left[\sum_{i=1}^2\bar{\phi}_i(-\Box+m^2_{0})\phi_i+\frac{\lambda}{4}(\bar{\phi}_1\phi_1)(\bar{\phi}_2\phi_2)\right].
\label{e:actionphiphi}
}
Here and in the following, $\mathbb{Z}_M$ projections on fields are written implicitly. 
%symbols to denote the fields such as $\phi_1$ implicitly means a projected fields $\hat{P}\phi_1$. 
Each vertex contribution involves three configurations of twists as mentioned in Fig.\ref{f:verttwist}. 
It is now almost clear that 
each twist of a vertex in $s$, $t$, and $u$-channels 
can be regarded as a twist of the propagator of the corresponding auxiliary field. 
 With the auxiliary field, the action has %The field has 
a three-point interaction vertex and reproduces the original four-point one when integrated out.

%In a correspondence
Corresponding to the above three ways for the opening of vertices, 
we can rewrite the action Eq.(\ref{e:actionphiphi}) into the following three forms:
\aln{
I_s&=\int\frac{d^{d+1}x}{M}\left[\sum_{i=1}^2\bar{\phi}_i(-\Box+m^2_{0})\phi_i+c_1c_2+i\frac{\sqrt{\lambda}}{2}c_1(\bar{\phi}_2\phi_2)+i\frac{\sqrt{\lambda}}{2}c_2(\bar{\phi}_1\phi_1)\right],
\label{e:actions}\\
I_t&=\int\frac{d^{d+1}x}{M}\left[\sum_{i=1}^2\bar{\phi}_i(-\Box+m^2_{0})\phi_i+\bar{d}d+i\frac{\sqrt{\lambda}}{2}\bar{d}\phi_1\phi_2+i\frac{\sqrt{\lambda}}{2}d\bar{\phi}_1\bar{\phi}_2\right],
\label{e:actiont}\\
I_u&=\int\frac{d^{d+1}x}{M}\left[\sum_{i=1}^2\bar{\phi}_i(-\Box+m^2_{0})\phi_i+\bar{d'}d'+i\frac{\sqrt{\lambda}}{2}\bar{d'}\,\bar{\phi}_1\phi_2+i\frac{\sqrt{\lambda}}{2}d'\,\bar{\phi}_2\,\phi_1\right].
\label{e:actionu}
} 
We have introduced three pairs of auxiliary fields: real scalars $(c_1, c_2)$, and complex scalars $(d, \bar{d}), (d', \bar{d'})$.\footnote{%---------------------------
The path integral contour for them should be chosen so that %as to make 
the partition function is convergent and thus the apparent violation of the reality or boundedness %boundness
in the above actions does not produce pathology.
} %-------------------------------------
Of course, each of Eqs.(\ref{e:actions})-(\ref{e:actionu}) is equivalent to Eq.(\ref{e:actionphiphi}) after integrating the auxiliary fields out.
Consequently, if we sum up the bubble diagrams from  all three models, 
we will encounter an overcounting at the level of free energy. 
However, when we consider configurations of twists, 
there is a one-to-one correspondence between vertex contributions of three channels and 
propagator contributions of each auxiliary field in these three models. 
In this sense, as far as a single twist of vertices is concerned, 
the vertex contributions we consider can be regarded as 
the propagator contributions from these three auxiliary fields.
As in Fig.\ref{Fig5}, 
four-point vertex contributions to EE with the  flux configurations, 
% $(m,0,-m,0)$, $(0,m,0,-m)$ or $(m,-m,m,-m)$
$(0,m,0)$, $(m,0,-m)$, or $(m,-m,m)$,  corresponds 
 to a propagator contribution of the associated auxiliary fields
 given by  Eqs.(\ref{e:actions})-(\ref{e:actionu}), respectively, for any bubble diagrams of the action Eq.(\ref{e:actionphiphi}).

\begin{figure}[t]
\centering
\includegraphics[width=10cm]{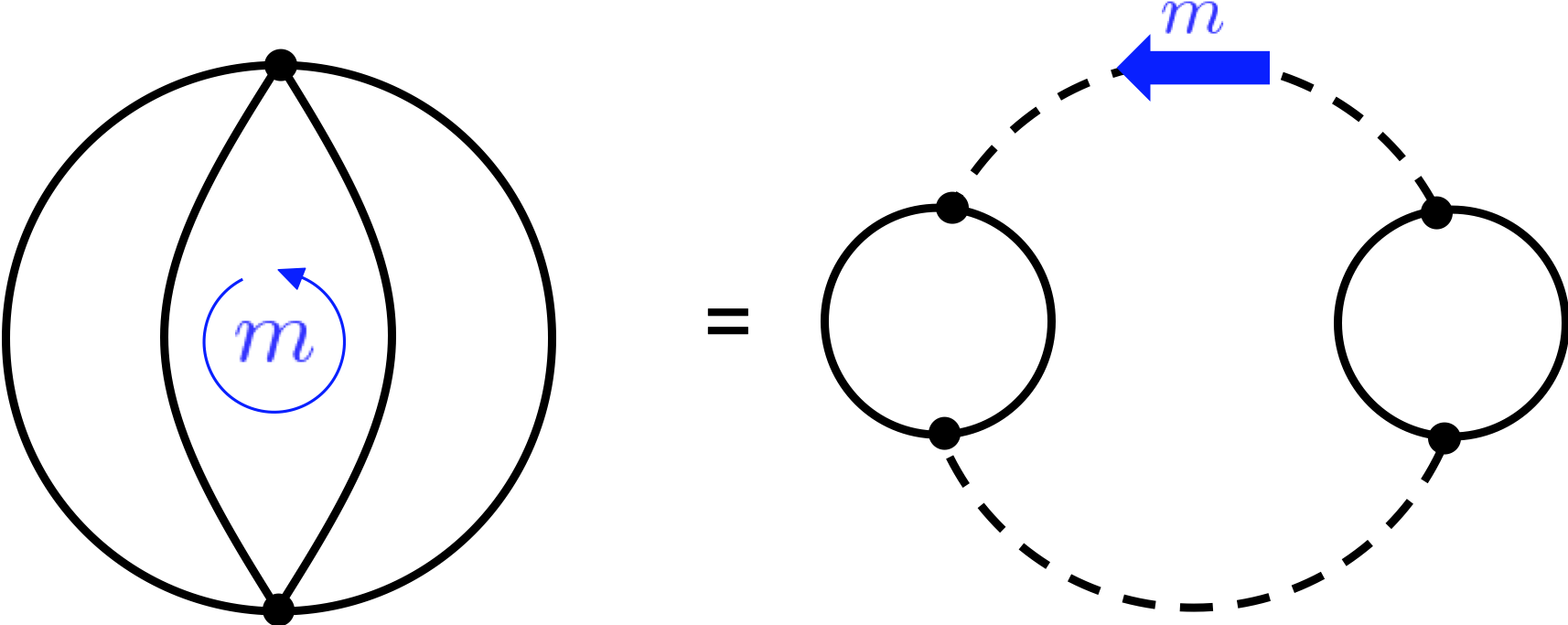}%{ambiguity2_ver2.pdf}
\caption{The right non-2PI diagram is obtained by opening two vertices in the left in terms of the auxiliary fields.
%An example of configurations with the one-to-one correspondence problem
%for a twist of a vertex. 
We can regard a flux of the center plaquette
as a twist of either upper or lower vertex, but not both.  
In terms of the auxiliary field,  
it is nothing but the phenomena  explained in Fig.\ref{f:ambiguity1}.
% in the non-2PI diagrams (right). 
}
\label{f:ambiguity2}
\end{figure}

\vspace{5mm}
Every vertex in the bubbles generated by Eq.(\ref{e:actionphiphi}) gets the contributions 
from the three channels\footnote{Figure-eight diagram is an exception and there is no $s$-channel.}.  
They coincide respectively with the contributions from a twisted propagator in the equivalent diagrams generated either by Eqs.(\ref{e:actions})-(\ref{e:actionu}). 
Here we have the same problem of the one-to-one correspondence 
between fluxes of twists in the plaquettes and twists of vertices, 
as mentioned in the previous subsection (Fig.\ref{Fig5}). 
In terms of the auxiliary fields, this problem is easily resolved by using the same logic
as in the propagator contributions. 
2PI diagrams do not have this kind of problem, and only 1-loop diagrams of the auxiliary fields
need care. See Fig.{\ref{f:ambiguity2}} as an example. 
As a result, the problem  is translated into the same problem for the twisted propagator of the auxiliary field. 
%%%%%%%%%%%%%%%%%%%%%%%%%%%%%%%%%%%%%%%%%%  
%%%%%%%%%%%%%%%%%%%%%%%%%%%%%%%%%%%%%%%%%%

\vspace{5mm}
The above observation leads us to express EE in the 2PI formalism with the auxiliary fields. 
Although we cannot rewrite the action itself by using all the auxiliary fields simultaneously, 
the vertex contributions to the free energy and EE can be written as a %direct 
sum of the contributions from these three. 
%It is because of the one-to-one correspondence explained above. 
The result is given by\footnote{%------------------------------------
Diagrams with tadpoles (one-point functions)  are cancelled  due to to the equation of motion. 
Namely, in calculating the 1PI free energy, 
 an appropriate source term is introduced 
depending on $M$ so that the equation of motion is always satisfied. 
} %--------------------------------------
\aln{
S^{\text{2PI}}_{\text{vert}}=&- \frac{V_{d-1} }{12} 
 \biggl(\int^{1/\epsilon} \frac{d^{d-1}k_{\parallel}} {(2 \pi)^{d-1}}  
 \tr \log \left[\tilde{G}_{c}^{-1}(\bm{0}; k_\parallel)   \right]
  \nonumber\\
 %&\hspace{1.6cm}
 &+2\int^{1/\epsilon} \frac{d^{d-1}k_{\parallel}} {(2 \pi)^{d-1}}  
  \log \left[ \tilde{G}_d^{-1}(\bm{0}; k_\parallel) \right] %\nonumber\\
% & \hspace{1.6cm}
 +2\int^{1/\epsilon} \frac{d^{d-1}k_{\parallel}} {(2 \pi)^{d-1}}  
  \log \left[ \tilde{G}_{d^\prime}^{-1}(\bm{0}; k_\parallel) \right]\biggr).
\label{e:EE2PIvertC}
}
Here, $(\tilde{G}_{c})_{ij}$, $\tilde{G}_d$ and $\tilde{G}_{d'}$ is the Fourier transformations of the two-point functions $\left<c_i(x)c_j(y)\right>$, $\left<d(x)\bar{d}(y)\right>$, and $\left<d'(x)\bar{d'}(y)\right>$ %, respectively. 
and the first, second, and third terms in Eq.\eqref{e:EE2PIvertC} 
represent the vertex contributions from the $s$-, $t$- and $u$-channel openings, respectively. The coefficients ``2'' in the second and third lines come from the fact that $(d,\bar{d})$ and $(d', \bar{d'})$ are complex fields. 
$(c_1,c_2)$ are real fields, but its propagator is written as a $2\times2$ matrix and has two degrees of freedom.  
The $\tr$ is the trace taken over this $2\times2$ matrix.

Eq.(\ref{e:EE2PIvertC}) has a remarkable interpretation. Note that we can regard the auxiliary fields as degrees of freedom of composite operators:
\aln{
c_1\sim \bar{\phi}_2\phi_2,~&~~c_2\sim \bar{\phi}_1\phi_1,\\
d\sim\bar{\phi}_1\bar{\phi}_2,~&~~\bar{d}\sim\phi_1\phi_2,\\
d'\sim\bar{\phi}_2\,\phi_1,~&~~\bar{d'}\sim\bar{\phi}_1\phi_2.
}
They are justified in various ways,  for instance, the vacuum expectation values of both sides coincide. From this viewpoint, Eq.(\ref{e:EE2PIvertC}) indicates that the vertex contributions are in fact understood as  propagator contributions of the composite operators. 
 From the actions Eqs.(\ref{e:actions}), (\ref{e:actiont}), and  (\ref{e:actionu}), the propagators of auxiliary fields are 
written in terms of correlation functions of the above composite operators as
\aln{
\tilde{G}_{cij}  &=(\sigma_x)_{ij} -\frac{\lambda}{4} \tilde{G}_s(\bm{k},k_\parallel)_{ij},
\\
\tilde{G}_{d} &=1 -\frac{\lambda}{4} \tilde{G}_t(\bm{k},k_\parallel),
\\
\tilde{G}_{d^\prime} &=1- \frac{\lambda}{4} \tilde{G}_u(\bm{k},k_\parallel),
}
where $\sigma_x$ is an $x$-component of the Pauli matrix and
\aln{
\tilde{G}_s(\bm{k},k_\parallel)_{ij}&=\int d^2\bm{r}\,d^{d-1}r_\parallel\,e^{-i(\bm{k}\cdot\bm{r}+ik_\parallel\cdot r_\parallel)}\left<[\bar{\phi}_j \phi_j](\bm{r};r_\parallel)~[\bar{\phi}_i\phi_i](\bm{0};0)\right>,\\
\tilde{G}_t(\bm{k},k_\parallel)&=\int d^2\bm{r}\,d^{d-1}r_\parallel\,e^{-i(\bm{k}\cdot\bm{r}+i k_\parallel\cdot r_\parallel)}\left<[\bar{\phi}_1\bar{\phi}_2](\bm{r};r_\parallel)~[\phi_1\phi_2](\bm{0};0)\right>,\\
\tilde{G}_u(\bm{k},k_\parallel)&=\int d^2\bm{r}\,d^{d-1}r_\parallel\,e^{-i(\bm{k}\cdot\bm{r}+ik_\parallel\cdot r_\parallel)}\left<[\bar{\phi}_2\phi_1](\bm{r};r_\parallel)~[\bar{\phi}_1\phi_2](\bm{0};0)\right>.
}
Thus the resulting contributions to EE, including both of those from the propagators and vertices, are given by 
\aln{
S^{\text{2PI}}_{\text{prop\&vert}}=&- \frac{V_{d-1} }{6} 
 \biggl( \sum_{i=1}^2\int^{1/\epsilon} \frac{d^{d-1}k_{\parallel}} {(2 \pi)^{d-1}}  
  \log \left[ \tilde{G_{\phi_i}}^{-1}(\bm{0}; k_\parallel) \epsilon^2 \right]
  \nonumber\\
 & \hspace{1.6cm}
 {-} \frac{1}{2} \int^{1/\epsilon} \frac{d^{d-1}k_{\parallel}} {(2 \pi)^{d-1}}  
 \tr  \log \left[ \sigma_x -\frac{\lambda}{4}\tilde{G_s}(\bm{0}; k_\parallel) \right]
  \nonumber\\
 & \hspace{1.6cm}
{-} \phantom{\frac{1}{2}} 
 \int^{1/\epsilon} \frac{d^{d-1}k_{\parallel}} {(2 \pi)^{d-1}}  
  \log \left[ 1-\frac{\lambda}{4} \tilde{G_t}(\bm{0}; k_\parallel)  \right]
  \nonumber\\
 & \hspace{1.6cm}
{-} \phantom{\frac{1}{2}} 
 \int^{1/\epsilon} \frac{d^{d-1}k_{\parallel}} {(2 \pi)^{d-1}}  
  \log  \left[1-\frac{\lambda}{4}\tilde{G_{u}}(\bm{0}; k_\parallel) \right] \biggr) ,
\label{e:EE2PIC}
} 
where the $\tr$ in the second line is a trace over the $ 2 \times 2$ matrix.

\vspace{5mm}
The above model is simple in the sense that the auxiliary field of each $s$, $t$, and $u$-channel is different and 
the correspondence between twisting a vertex and twisting  propagator  of each auxiliary field is clear. 
Let us then consider a less easy (though seemingly easier) case, namely the $\phi^4$-theory with a single real scalar. 
The action written with an auxiliary field $c$  takes the following form:
\aln{
I_{stu}=\int\frac{d^{d+1}x}{M}\left[\frac{1}{2}\phi(-\Box+m^2_{0})\phi+\frac{1}{2}c^2+i\sqrt{\frac{\lambda}{2}}\,c\,\phi^2\right]. 
\label{e:actionc}
}
In order to reproduce the vertex contributions to EE in the 
original  $\lambda\phi^4/4$ theory, we need to sum all the contributions from the three different channels for $c$. 
If we use the above action,  the free energy in flat space can be reproduced, 
but not the free energy of the orbifold theory. 
Thus we cannot use the renormalized two-point function of $c$ via $\log G_c^{-1}$ to express the correct
amount of vertex contributions to EE. 
EE in $\phi^4$ theory 
 is neither expressed by a single auxiliary field $c$ nor by triple copies of it
because the three channels coincide and  get mixed among them. 

In spite of this difficulty, 
 we can still get a consistent description of vertex  contributions, 
	 not through the auxiliary field method, but directly in terms of the composite operator. 
	As the previous observation indicates, we will now focus on the following correlation function,
	\aln{
		G_{stu}(x-y):=\left<:\!\phi^2\!:\!(x) :\!\phi^2\!:\!(y)\right>.
	}
	The vertex contributions to EE in the $\phi^4$ theory is expected to be given by 
	\aln{
		S^{\text{2PI}}_{\text{vert}}=
		\frac{V_{d-1}}{12}
		\int^{1/\epsilon}\frac{d^{d-1}k_\parallel}{(2\pi)^{d-1}}\mathrm{log}\,\left[1-\frac{3}{2}\lambda \,\tilde{G}_{stu}(\bm{0},k_\parallel)
		\right] .
		\label{e:EEcomp}
	}
	Here, the coefficient $-3\lambda/2$ is understood as $-\lambda/4\times 6$ %the former of which is 
	where $-\lambda/4$ is the coefficient in front of the interaction vertex (the same coefficient as in Eq.(\ref{e:EE2PIC})) 
	and the coefficient 6 is the combinatorial factor for separating four $\phi(x)$'s into a pair of  two $\phi(y)$'s. 
	The unity in the logarithm in Eq.(\ref{e:EEcomp}) means that 
	the composite operator does not have any new degrees of freedom  in the free field limit and does not contribute to EE. 
	The overall factor is not $1/6$ but $1/12$ since the composite operator is real. 
%	To summarize, We expect that Eq.(\ref{e:EEcomp}) is a consistent counterpart to Eq.(\ref{e:EE2PIvertC}) (or equivalently, the second term and the followers in Eq.(\ref{e:EE2PIC})), which describes the vertex correction in the $\phi^4$ theory and $1-(3\lambda/2)G_{stu}$ should correspond to the full propagator of the necessary degrees of freedom for opening vertices, even though we cannot equip it at the level of action. 

Since we cannot introduce the auxiliary field and use the conventional  2PI formalism, 
we do not yet know how to prove that the above expression of Eq.(\ref{e:EEcomp}) gives the correct vertex contributions to EE.
Instead, we will perturbatively check its correctness up to  $\lambda^2$ in the following. 
The two-point function of the composite operator can be evaluated as 
	\aln{
		G_{stu}=2A-6\lambda A^2-12\lambda B+O(\lambda^2),
		\label{e:comppropexpand}
	}  
	where
	\aln{
		A &:=G_{0}(x-y)^2, \nn
		B& :=\int d^{d+1}z\,G_{0}(x-y)G_{0}(x-z)G_{0}(z-y)G_{0}(0).
	}
	In Eq.(\ref{e:comppropexpand}),  
	the product of operators represents a convolution; $XY(x-y)=\int d^{d+1}z X(x-z)Y(z-y)$. 
	By substituting Eq.(\ref{e:comppropexpand}) into Eq.(\ref{e:EEcomp}), and using the identity
	\aln{
		\int\frac{d^{d-1}k_\parallel}{(2\pi)^{d-1}}\tilde{f}(\bm{0},k_\parallel)=\int d^2\bm{r}f(\bm{r},0),
	}
	we can expand Eq.(\ref{e:EEcomp}) up to $O(\lambda^2)$ as 
	\aln{
		S^{\text{2PI}}_{\text{vert}}&=\frac{V_{d-1}}{12}\int d^{2}\bm{r}\left[\log\,\left(1-\frac{3}{2}\lambda G_{stu}\right)\right](\bm{r},0)\nonumber\\
		&=\frac{V_{d-1}}{12}\int d^{2}\bm{r}\left[-\frac{3}{2}\lambda G_{stu}-\frac{9}{8}\lambda^2 G_{stu}^2\right](\bm{r},0)
		+ {\cal O}(\lambda^3)
		\nonumber\\
		&=\frac{V_{d-1}}{12}\int d^{2}\bm{r}\left[-3\lambda A+18\lambda^2B+\frac{9}{2}\lambda^2 A^2\right](\bm{r},0)
		+ {\cal O}(\lambda^3) \nonumber\\
		&=-\frac{V_{d-1}}{4}\lambda\int d^2\bm{r}\,G_{0}(\bm{r},0)^2\nonumber\\
		&~~+\frac{3V_{d-1}}{2}\lambda^2\int d^2\bm{r}\,d^2\bm{s}\,d^{d-1}r_\parallel\, G_0(\bm{r},0)\, G_0(\bm{r}-\bm{s},r_\parallel)\,G_0(\bm{s},r_\parallel)G(\bm{0},0)\nonumber\\
		&~~+\frac{3V_{d-1}}{8}\lambda^2\int d^2\bm{r}\,d^2\bm{s}\,d^{d-1}r_\parallel\, G_{0}(\bm{r},r_\parallel)^2\,G_0(\bm{s},r_\parallel)^2
+ {\cal O}(\lambda^3)	.}
	These three terms indeed coincide with Eqs.(\ref{e:EE2loopvert}), (\ref{e:EE3loopvert2}), and (\ref{e:EE3loopvert}), respectively.

The present result is surprising, or rather amusing 
since   the non-Gaussian contributions to EE can be 
 understood in terms of two-point functions of composite operators
even when the auxiliary field can not be consistently introduced. 
As explained in Sec.\ref{sec:twist}, a twisted propagator is pinned with loose ends
reflecting quantum correlations between two spacial regions. 
From this observation, it is tempting to expect that EE can be  interpreted as a sum of 
correlations of various composite operators, not restricted to those that appear at the classical action. 
Indeed, in the framework of the Wilsonian RG, the effective action (EA) changes as the 
energy scale is changed, and the EA contains infinitely many vertices. 
Thus EE will also follow the same RG flow. We want to come back to this important issue in near future.  
%Note that they are pinned at the boundary and hence measure quantum correlations across the boundary
%though composite operators. 

%%%%%%%%%%%%%%%%%%%%%%%%%%%%%%%%%%%%%%%%%%%%%%%%%%%%%
 \section{Generalizations to theories with spins} 
\label{s:spin}
%%%%%%%%%%%%%%%%%%%%%%%%%%%%%%%%%%%%%%%%%%%%%%%%%%%%%
All the above studies have been focused on scalar field theories. 
The analysis can be easily extended to a scalar theory with multiple flavors. 
Furthermore, we can straightforwardly extend it to general field theories with spins. 
As briefly explained at the end of Sec.\ref{s:orbifold}, we need an additional phase rotation
corresponding to its spin. 
Besides a modification necessary for fermionic fields and subtlety for higher spin fields of $s \ge 3/2$, 
 the orbifold method is applicable to them. 

A twisted propagator with a spin-$s$ field $\varphi_s$ is accompanied with a rotation in the internal space:
\aln{
G_{\varphi_s\,0}^{(M)}(x,y)=\sum_{m=0}^{M-1}e^{-2i\theta_m\mathcal{M}^{(s)}_{1,d+1}}\,G_{\varphi_s\,0}(\hat{g}^mx-y). 
}
Here, $\mathcal{M}_{1,d+1}^{(s)}$ is one of the generators of $SO(d+1)$ in the spin-$s$ representation, which drives a rotation on a plane spanned by $x_\perp$ (1-direction) and $\tau$ ($(d+1)$-direction). For example, the propagator for a Dirac fermion is given by
\aln{
S^{(M)}(x,y)=\sum_{m=0}^{M-1}e^{\theta_m \gamma_1\gamma_{d+1}}\int\frac{d^2\bm{k}}{(2\pi)^2}\frac{d^{d-1}k_\parallel}{(2\pi)^{d-1}}\frac{i\bm{k}\cdot\bm{\gamma}+ik_\parallel\cdot\gamma_\parallel-m_{0}}{k^2+m^2_{0}}e^{i(\bm{k}\cdot\hat{g}^m\bm{x}-\bm{k}\cdot\bm{y}+k_\parallel\cdot(x_\parallel-y_\parallel))}
}%Note that 
%If 
with $\bm{\gamma}=(\gamma_1,\gamma_{d+1})$ and $\gamma_\parallel=(\gamma_2,\cdots,\gamma_{d-2})$. 

In a bubble diagram, each propagator has such an additional rotational factor. However, since an interaction vertex is rotationally invariant, 
it is still invariant under $\mathbb{Z}_M$ rotation and consequently invariant under
 an overall twist of the adjacent propagators.\footnote{%---------------------------------
A simple example is a vertex in the $U(1)$ gauge theory, $(\gamma_\mu)_{\alpha\beta}$. It has one vector field and two spinor fields
 and is invariant under simultaneous rotations of the fields. 
} %---------------------- 
Suppose that we have a multi-point vertex of fields with spins $s_q$ ($q=1,2, \cdots$) and the coefficient is given by $C_{i_1i_2\cdots}$.
The $\mathbb{Z}_M$ invariance of the vertex is written as 
\aln{
C_{i_1i_2\cdots}\delta^2(\bm{p}_1+\bm{p}_2\cdots)&=(e^{2i\theta_m \mathcal{M}_{1,d+1}^{(s_1)}})_{i_1}^{~j_1}(e^{2i\theta_m \mathcal{M}_{1,d+1}^{(s_2)}})_{i_2}^{~j_2}\cdots C_{j_1j_2\cdots}\delta^2(\hat{g}^{m}(\bm{p}_1+\bm{p}_2\cdots)).
}
By decomposing each field  into irreducible representations of $SO(2)$, this simply means that  a sum of $SO(2)$ spins vanish
at each vertex. Due to the invariance, the basic framework of $\mathbb{Z}_M$ gauge theory on Feynman diagrams is not
changed. Namely, we can classify $\mathbb{Z}_M$ invariant configurations of twists in terms of  fluxes in plaquettes as before.
The additional phase associated with spins can be calculated by taking a special gauge of $\mathbb{Z}_M$ fluxes because of their gauge invariance.

Another  point to notice 
is that,  for fermions, we have to replace the twist operator $\hat{g}$ with $\hat{g}^2$
due to the anti-periodic boundary condition. 
In this case, $M$ should be considered as an odd integer. 

\medskip

In presence of higher spin fields, we can repeat the 2PI analysis. 
As far as the contributions from the propagators and vertices are concerned, 
it is sufficient to consider a  twist of a particular propagator (or composite operator), and 
the additional phase can be easily obtained. 
 For a general bosonic or fermionic field $\varphi_s$ with spin $s$, we can formally write down the free energy:
\aln{
\tilde{F}^{\text{2PI}}_{\varphi_s,\text{prop}}&=\frac{V_{d-1}}{2M}\sum_{m=1}^{M-1}\frac{1}{4\sin^2\theta_m}\mathrm{tr}\left[e^{2i\theta_m \mathcal{M}^{(s)}_{1,d+1}}\,\int\frac{d^{d-1}k_\parallel}{(2\pi)^{d-1}}\mathrm{log}\,\tilde{G}_{\varphi_s}(\bm{0}, k_\parallel)\right]~~(\text{for bosons}),
\label{e:freeenergyspinbos}\\
\tilde{F}^{\text{2PI}}_{\varphi_s,\text{prop}}&=-\frac{V_{d-1}}{2M}\sum_{m=1}^{M-1}\frac{1}{4\sin^22\theta_m}\mathrm{tr}\left[e^{4i\theta_m \mathcal{M}^{(s)}_{1,d+1}}\,\int\frac{d^{d-1}k_\parallel}{(2\pi)^{d-1}}\mathrm{log}\,\tilde{G}_{\varphi_s}(\bm{0}, k_\parallel)\right]~~(\text{for fermions}).
\label{e:freeenergyspinfer}
}
``$\mathrm{tr}$'' here represents the trace over the internal space. Moreover, the vertex contributions are written in terms of the renormalized propagators of composite operators as well as the scalar field case. When one considers a general composite operator such as $:\!\varphi_s\varphi'_{s'}\!:$, it is generically in a reducible representation of $SO(d+1)$. 
We first decompose it into irreducible components, each of which corresponds to a different composite operator.

If we reduce Eqs.(\ref{e:freeenergyspinbos}), (\ref{e:freeenergyspinfer}) to the free field cases, we can easily evaluate the trace because both the rotational factor and $\tilde{G}_{\varphi\,0}(\bm{0};k_\parallel)$ are diagonalized in the basis of the eigenstates for $SO(2)$. 
The resulting EEs coincide with those in \cite{He:2014gva}. 
On the other hand, for interacting cases, $\tilde{G}_{\varphi}(\bm{0};k_\parallel)$ has off-diagonal components 
 and we need to take a trace  of {the product of the rotational factor and the matrix-valued logarithmic terms} in a nontrivial way.
It is technically difficult to proceed to  further computations and we leave it for future investigations.
Meanwhile, we can conclude that the non-Gaussian part in EE is understood as contributions from renormalized two-point functions of composite operators while the Gaussian part is  a contribution from the fundamental  %and composite 
fields.

%%%%%%%%%%%%%%%%%%%%%%%%%%%%%%%%%%%%%%%%%%%%%%%%%%%%%
 \section{Conclusions and discussions}
\label{s:discussion}
%%%%%%%%%%%%%%%%%%%%%%%%%%%%%%%%%%%%%%%%%%%%%%%%%%%%%
In the present paper, we have studied EE in general interacting QFTs %in general 
from the field theoretical {perspective} proposed in our previous work \cite{Iso:2021vrk}.
The approach is based on the orbifold method to calculate EE of half space and the
consequent idea of $\mathbb{Z}_M$ gauge theory on Feynman diagrams. 
In this method, EE is given by a sum of various configurations of $\mathbb{Z}_M$ fluxes on each
of the plaquettes  in Feynman diagrams. 
Among infinitely many configurations of fluxes, we have extracted two dominant 
contributions to EE, that correspond to twisting propagators and vertices. 
An essential development in the present paper 
from our previous work \cite{Iso:2021vrk} is a new interpretation 
of the vertex contributions in terms of correlation functions of composite operators.  
We have also shown that the propagator contributions to EE are exactly given by the full renormalized propagators
in the 2PI formalism
 {where} two-point functions are treated nonperturbatively, and 
 as a consequence, we have succeeded to fully extract the Gaussian contributions to EE, %and 
Thus the vertex contributions that are interpreted as correlations of composite operators
 purely represent the non-Gaussianity of the vacuum. 

Then, {one of} the most crucial questions left {unanswered} is how we %should
{can} understand or evaluate configurations of twists other than those corresponding to a single twisted propagator or vertex. 
We {might} %will 
be able to address this question by developing an efficient computational method in the $\mathbb{Z}_M$ gauge theory on Feynman diagrams. On the other hand, we may be able to extract further contributions by  extending our finding
that some of the non-Gaussian contributions are interpreted as correlations of composite operators. 
Suppose that a bubble diagram is separated into two pieces connected by a “fat” propagator.
Then, 
twisting the fat propagator might give a contribution to EE associated with the correlation of the macroscopic composite operators. 
It is tempting to expect that 
a general configuration of twists would be understood as a correlation of various 
composite operators between two spatial regions. 
If this expectation is true, then what quantity of composite operators will give  the magnitude of its contributions
to EE? A naive guess is its mass dimension or the correlation length. 
In our studies, we have extracted contributions of the propagator and the vertices. 
The propagator contribution to EE is given by  scalar fields with the mass dimension 1, 
while the vertex contributions are given by composite operators with the mass dimension 2. 
It is the reason why we think that they give dominant contributions to EE compared to others.

We may also apply the method of Wilsonian RG to extract further contributions to EE.
In the present paper, we have considered such vertex contributions that the vertices are 
already present in the classical action. 
In the Wilsonian RG picture, the effective action is scale-dependent
and contains many other vertices besides those present in the classical action. 
Then we may introduce further auxiliary fields corresponding to various composite operators 
whose mass dimensions are higher than 2. 
Since these composite operators are expected to decay faster than those studied in the present paper,
their contributions will be less dominant but we may be able to extract 
contributions to EE systematically by using the Wilsonian effective action. 

		Along the Wilsonian RG flow, quantum vertices appear in addition to classical vertices while the dynamical degrees of freedom to be integrated decrease. From this perspective, we can partly answer the unsolved questions: Do the other configurations of twists not discussed in Sec. \ref{s:prop} and \ref{s:vert} contribute? For instance, a general flux configuration of the figure-eight 2-loop diagram is characterized by two $\mathbb{Z}_M$ integers while only $(m,0)$, $(0,m)$, and $(m,-m)$ can be identified as twists of propagators or vertices (Fig.\ref{Fig3}). The fluxes other than these special sets should be regarded as twisting more than one propagator or vertex and cannot be attributed to a single propagator or vertex. Thus we could not evaluate such contributions. However, this problem at the original UV theory is no more a problem in the IR effective theory.
		After renormalization, such configurations of fluxes in the IR limit are either interpreted as twisting general vertices, which implies twisting more general composite operators, or abandoned as the UV part of contributions and absent in the IR universal part. They should be treated in the Wilsonian effective field theory and now under investigation~\cite{IMS3}.
	
Furthermore,  the Wilsonian RG approach will  tell us how we should take the renormalization scale in
the calculation of EE and also how EE varies along the RG flow. 
In relation to this, it is worthwhile to investigate 
how our results are connected to another method using
the continuous multi-scale entanglement renormalization ansatz \cite{Fernandez-Melgarejo:2020utg,Fernandez-Melgarejo:2021ymz,Fernandez-Melgarejo:2020fzw}. 

It is also interesting to generalize our results 
 to other {choices of} spatial subregions. 
Our investigation depends heavily on the orbifold method, which is not applicable for regions other than the half space of flat space. However, the results and observations have  general implications based on two-point functions pinned at the boundary. 
As we have shown, twisted propagators are pinned at the boundary, which can be 
interpreted as reducing the degrees of freedom in the normal direction to the boundary. 
If the boundary has a more nontrivial shape, we need a clever way to foliate the space
to specify correlations between such two spacial regions. 
It is tempting to relate it to 
the holographic view of EE \cite{Ryu:2006bv,Ryu:2006ef,Hubeny:2007xt}, or a geometric perturbation  \cite{Rosenhaus:2014zza}.

Other generalizations include EE in non-relativistic QFTs or EE of excited states.
Since our approach uses a standard QFT technique of Feynman diagrams, it should be relatively easy to study EE of excited states. 
For example, EE of an excited state by operators $O$ must be obtained by replacing the free energy with the corresponding correlation functions $\langle O_1^\dagger O_1 \cdots O_M^\dagger O_M \rangle$, where $O_i$ is given by $O$ restricted within the half of the $i$-th piece of the orbifolded space. In CFTs, some concrete calculations based on this approach are performed and
applied to the orbifold method~\cite{Caputa:2017tju} as well as the standard replica trick~\cite{Nozaki:2014hna,Nozaki:2014uaa,Caputa:2014vaa,Nozaki:2015mca,Chen:2015usa,He:2014mwa,Caputa:2015tua,Numasawa:2016kmo,Nozaki:2016mcy}.
As for a practical application, we could explicitly investigate the entropic c-theorem \cite{Casini:2004bw} along the RG flow based on the direct computation of EE in the presence of interactions.

\begin{acknowledgments}
%{\it Acknowledgments} ---
%We thank Yoshiki Sato, Sotaro Sugishita, Takao Suyama and Tadashi Takayanagi for valuable comments. 
We are supported in part by the Grant-in-Aid for Scientific research, No. 18H03708 (S.I.), No. 16H06490 (S.I.),  No. 20J00079 (K.S.)
and SOKENDAI. 
\end{acknowledgments}

%\begin{thebibliography}{99}
%\end{thebibliography}

\bibliographystyle{apsrev4-1}
\bibliography{EE}
%\bibliography{ent,ent2}
\end{document}